\documentclass[12pt,preprint]{aastex}
\usepackage{epsfig}
\usepackage{lscape}
\usepackage{rotating}
\usepackage{natbib}

\newcommand{\xmm}{{\textit XMM}}
\newcommand{\xmmn}{{\textit XMM-Newton}}

\begin{document}
\title{Evidence for Black Hole Growth in Local Analogs to Lyman Break Galaxies}
\author{Jianjun Jia\altaffilmark{1}, Andrew Ptak\altaffilmark{2}, Timothy M. Heckman\altaffilmark{1}, Roderik A. Overzier\altaffilmark{3}, Ann Hornschemeier\altaffilmark{2}, Stephanie M. LaMassa\altaffilmark{1}}
\affil{\altaffilmark{1}Department of Physics and Astronomy, Johns Hopkins University, Baltimore, MD 21218, USA\\
\altaffilmark{2}Goddard Space Flight Center, Greenbelt, MD 20771, USA\\
\altaffilmark{3}Max-Planck Institute for Astrophysics, D-85748 Garching, Germany}

\begin{abstract}
We have used \xmmn\ to observe six Lyman Break Analogs (LBAs): members of the rare population of local galaxies that have properties that are very similar to distant Lyman Break Galaxies. Our six targets were specifically selected because they have optical emission-line properties that are intermediate between starbursts and Type 2 (obscured) AGN. Our new X-ray data provide an important diagnostic of the presence of an AGN. We find X-ray luminosities of order 10$^{42}$ erg s$^{-1}$ and ratios of X-ray to far-IR luminosities that are higher than values in pure starburst galaxies by factors ranging from $\sim$ 3 to 30. This strongly suggests the presence of an AGN in at least some of the galaxies. The ratios of the luminosities of the hard (2-10 keV) X-ray to [O III]$\lambda$5007 emission-line are low by about an order-of-magnitude compared to Type 1 AGN, but are consistent with the broad range seen in Type 2 AGN. Either the AGN hard X-rays are significantly obscured or the [O III] emission is dominated by the starburst. We searched for an iron emission line at $\sim$ 6.4 keV, which is a key feature of obscured AGN, but only detected emission at the $\sim$ 2$\sigma$ level. Finally, we find that the ratios of the mid-infrared (24$\mu m$) continuum to [O III]$\lambda$5007 luminosities in these LBAs are higher than the values for Type 2 AGN by an average of 0.8 dex. Combining all these clues, we conclude that an AGN is likely to be present, but that the bolometric luminosity is produced primarily by an intense starburst. If these black holes are radiating at the Eddington limit, their masses would lie in the range of 10$^5$ to 10$^6$ M$_{\odot}$. These objects may offer ideal local laboratories to investigate the processes by which black holes grew in the early universe.
\end{abstract}


\section{Introduction\label{introduction}}

One of the major unsolved problems in astrophysics today is the nature of the connection between the formation and evolution of galaxies and supermassive black holes (SMBHs). Current models invoke a close interplay  (``feedback") between SMBHs and their host galaxies in order to arrest the growth of the most massive galaxies, which are otherwise over-predicted by models of structure formation \citep[e.g.,][]{springel03,croton06}. 

Observations of high-redshift galaxies paint a somewhat confusing picture of the relationship between star formation and black hole growth \citep[e.g.,][]{heckman09}. The best-studied population of high-redshift star-forming galaxies are the Lyman-Break Galaxies (LBGs) and closely related types of UV-selected galaxies. They represent a major phase in the early stages of galaxy formation and evolution \citep{giavalisco02}. They are a primary tracer of high redshift star formation, as they are easily found at $z=2-6$ in deep pencil beam optical surveys \citep[e.g.,][]{steidel96,shapley03,giavalisco04}. The LBGs are believed to be the precursors of present day galaxies undergoing a phase of intense star formation and merging, and a large fraction of LBGs may merge to form early-type galaxies that are situated in present-day groups and clusters \citep[e.g.,][]{adelberger05}. 

Most of the LBGs do not harbor an Active Galactic Nucleus (AGN) luminous enough to be individually detected in the deepest X-ray surveys. However, a subset ($\sim$3\%) of LBGs with actively accreting SMBHs has been identified \citep{steidel02,laird06}. These AGNs typically have X-ray luminosities of $10^{42}-10^{43}$ erg s$^{-1}$ in the 2-10 keV energy range, and often show signs of significant obscuration. Detailed studies of what might trigger both the AGNs and the starbursts in these forming galaxies are difficult since their distances render them small and faint.

The Galaxy Evolution Explorer (\textit{GALEX}) mission is conducting surveys in the far- and near-UV ($\lambda \simeq$ 1600 and 2300 \AA~respectively) and has  been used to identify a rare population of low-redshift ($z\sim$0.1-0.3) galaxies that are remarkably similar to high-redshift LBGs in most physical properties \citep{heckman05a,hoopes07,overzier08, overzier09, overzier10}. This population was first identified by \cite{heckman05a} and \cite{hoopes07}, based on cross-matching the \textit{GALEX} and Sloan Digital Sky Survey data. These objects were defined by two criteria designed to match the typical LBGs: 1) a minimum luminosity in the far-UV (FUV) band $L_{FUV}\ge 2\times 10^{10}L_{\odot}$; 2) an FUV effective surface brightness $I_{FUV}\ge 10^{9}L_{\odot}\rm{kpc}^{-2}$, where $I_{FUV}$ is the mean FUV surface brightness interior to the SDSS $u$ half-light radius \footnote{While selection based on high FUV luminosity and surface brightness also selects Type 1 AGN and BL Lacs, these objects have been eliminated from samples of LBAs \citep[see][for details]{hoopes07}. Note also that \cite{overzier09} presented very high signal-to-noise optical spectra of LBAs showing that there is no spectroscopic signature of a Type 1 AGN.}. These local ``Lyman break analogs'' (LBA) are similar to distant LBGs in size, surface brightness, metallicity, dust, star formation rate (SFR), stellar mass, and gas velocity dispersion.  There are other local star forming galaxies, e.g., Ultra-Luminous Infrared Galaxies (ULIRGs) and Blue Compact Dwarfs (BCDs), which have some similarities with LBGs. However, ULIRGs are much more heavily obscured by dust, and BCDs are much smaller in mass and luminosity (star-formation rate) compared with LBGs. These differences make them poorer local analogs to LBGs.  However, it is intriguing that an actively accreting SMBH has recently been reported in a nearby dwarf starburst galaxy \citep{reines11}.

Interestingly, about 20\% of the LBAs have optical emission-line spectra that are intermediate between those of pure starbursts and those of Type 2 (obscured) AGN. As discussed by \cite{overzier09}, the optical spectra alone do not unambiguously establish the presence of a Type 2 AGN. They may instead be extreme starburst systems in which the physical conditions in the ionized gas are different from typical local starbursts.  

\cite{overzier09} also presented imaging with the \textit{Hubble Space Telescope} that showed that most of these possible starburst-AGN composites contain extremely massive (one to few-billion M$_{\odot}$) and compact (radius $\sim 10^2$ pc) starbursts. These Dominant Compact Objects (DCOs) would appear to be ideal sites for the formation and/or rapid growth of a SMBH. It has been suggested that the seeds of present-day SMBHs may be present at the centers of the massive, compact stellar clumps seen in clumpy galaxies at high redshift \citep[e.g.][]{elmegreen08}. If these LBAs indeed host an AGN, they would offer the opportunity to study the relationship between the formation of galaxies and supermassive black holes in detail in objects that appear to be excellent local analogs to a major population of high-redshift galaxies.  Unfortunately, the optical spectra alone do not unambiguously establish that these LBAs are indeed composite starburst/obscured-AGN systems (as opposed to starbursts with extreme properties) \citep{overzier09}.  Observations in the hard X-ray band can potentially offer us important new diagnostics of the presence of an obscured AGN.

In this paper we report the results of such an investigation. We describe our data in \S \ref{data} and present the spectral analysis in \S \ref{results}. The results are discussed in \S \ref{discussion}, and we come to some brief conclusions in \S \ref{conclusion}.

\section{Data Analysis\label{data}}
\subsection{Sample selection}
\begin{figure}
\centering
\epsfig{file=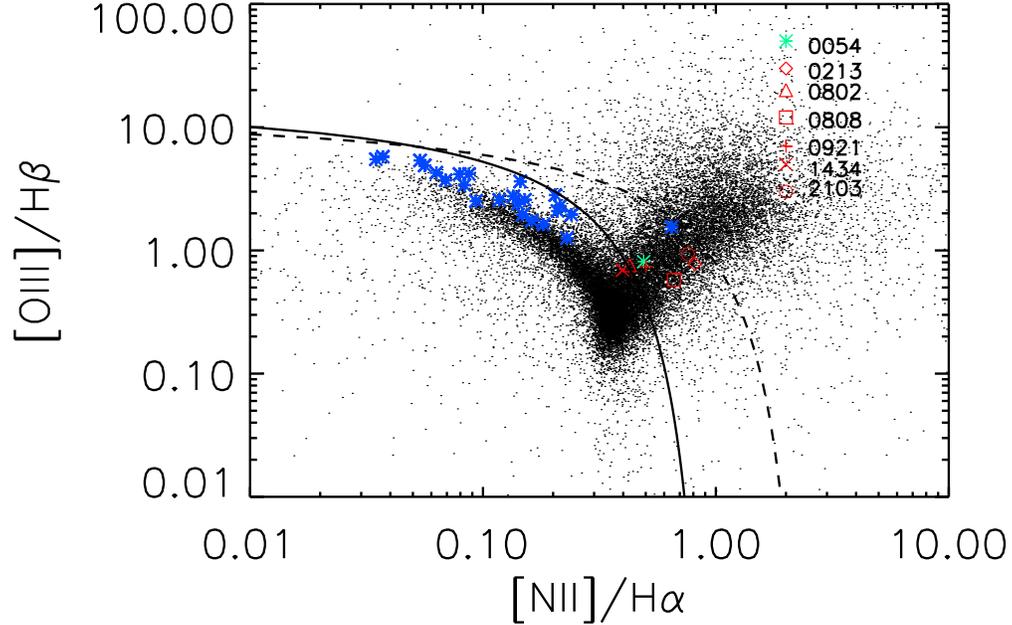,width=0.8\linewidth}
\caption{Optical emission line diagnostic diagram showing the 31 well-studied nearby ($z\le 0.3$) LBAs (colored symbols ) discovered by \protect\cite{heckman05a,hoopes07}. The LBAs marked in red in this paper lie in the star-forming/AGN composite region (we did not get observation time of the object marked in green), where the solid line is from \cite{kauffmann03}, and the dashed line is from \protect\cite{kewley06}. \label{f:bpt}}
\end{figure}

To select our sample for X-ray observations we begin with the 31 LBAs with \textit{HST} imaging \citep{overzier09} and \textit{Spitzer} (IRAC+MIPS) photometry \citep{overzier11}. In Fig. \ref{f:bpt} we show the standard ``BPT" diagram \citep{bpt} used to classify the optical emission-line spectra of galaxies, where the emission line ratios are from Table 2 in \cite{overzier09}. We plot all the galaxies in the \textit{GALEX}-SDSS cross-matched sample and then highlight the 31 LBAs in our sample (large symbols). We note that seven\footnote{We include SDSS J1434+0207 (red cross in Fig. \ref{f:bpt}) in our ``LBA-composite" sample, although it lies close to the lower left of the solid line in the BPT diagram and does not harbor a DCO.} of the LBAs lie in the composite region in the ``BPT" diagram, which is known to harbor ``composite objects", which are typically (but not necessarily) objects having both star-formation and a Type 2 AGN \citep[see][]{kauffmann03,kewley06}. Hereafter, we refer to this sample as ``LBA-composites". In this paper, a flat $\Lambda$ cold dark matter ($\Lambda$CDM) cosmology with $h=0.7$, $\Omega_{\Lambda}=0.7$ is assumed. 

\subsection{The X-ray data}
\xmmn~observations were performed for 6 of the 7 LBA-composites.  SDSS J0054+1554 was given a ``C" priority and not observed (marked as green asterisk in Fig. \ref{f:bpt}). A summary of observations is shown in Table \ref{t:summary}.

The reduction of the \xmmn~ data was done by \texttt{XAssist (ver. 0.97)}\footnote{http://xassist.pha.jhu.edu/zope/xassist}, which is a software for automatic analysis of X-ray data. \texttt{XAssist} runs the \xmmn~ Science Analysis System (\texttt{SAS}) packages automatically to filter the data, generate the light curves and extract the spectra of sources in the field, as well as compute the associated redistribution matrix files (RMF) and ancillary region files (ARF). The LBA-composites are weak in X-ray emission, and their optical positions were used to define the source regions in X-ray data reduction. \texttt{XAssist} determines the source extraction sizes for source by fitting an elliptical Gaussian model to the source image, which in the case of on-axis \xmm~observations typically results in regions of size 20-25''.  Background regions were chosen as annuli centered on the sources to extract the background spectra. Table \ref{t:cstat} lists the PN and MOS counts for each source.

\subsection{Ancillary data}
The Optical Monitor (OM) aboard on \xmmn~allows us to make simultaneous observations in optical/UV and X-ray bands. We processed the OM data with the \texttt{omichain} package in \texttt{SAS} to obtain UV count rates and the corresponding magnitudes of each object, as well as their associated uncertainties. We obtained exposures in the U (344nm), UVW1 (291nm), UVM2(231nm) and UVW2 (212nm) band filters, but only the UVW1 was used for all six LBAs. We list the UV detections with each filter and the luminosity measured in UVW1 in Table \ref{t:uv}. The conversions from UV count rates to specific fluxes (in units of erg/s/cm$^2$/\AA) were calculated based on the white dwarf standard stars\footnote{http://xmm.esa.int/sas/current/watchout/\\Evergreen\_tips\_and\_tricks/uvflux.shtml}.

Table \ref{t:loiii} lists the observed and extinction-corrected [O III]$\lambda$5007 line luminosity ($L_{\rm{[O III],obs}}$ and $L_{\rm{[O III],corr}}$, respectively) for each LBA-composite. We obtain $L_{\rm{[O III],obs}}$ from the SDSS Data Release 7 MPA/JHU catalog. We derive $L_{\rm{[O III],corr}}$ from $L_{\rm{[O III],obs}}$ using the observed H$\alpha$/H$\beta$ ratio, an intrinsic (H$\alpha$/H$\beta$)$_{0}$ ratio of 2.9, and the standard extinction curve for galactic dust \citep{osterbrock06}. In Table \ref{t:loiii}, we also list the mid-infrared (MIR) and far-infrared (FIR) luminosity, $L_{24\mu m}$ and $L_{\rm{FIR}}$, respectively \citep{overzier11}. 

\section{Spectral Analysis\label{results}}
\subsection{X-ray Spectra}
X-ray spectral fitting was performed with \texttt{XSPEC} (ver. 12).  X-ray photons are collected by three detectors on \xmmn, i.e. PN, MOS1 and MOS2. The data of three detectors can be fitted simultaneously in \texttt{XSPEC} by tying all parameters together except a multiplicative constant factor. We froze this factor at 1 for PN, and allowed it to vary between 0.8 to 1.2 for MOS1 and MOS2. Since our targets have limited counting statistics (see Table \ref{t:cstat}), we first fit their spectra using the $C$-statistic \citep{cash79}, which is often used in this low counts regime. The spectra were grouped to 1 count per bin, to improve the performance of \texttt{XSPEC} \citep{teng05}.  The simplest spectral model often fit to AGN is an absorbed power-law model. The X-ray photons suffer from two sources of absorption.  The first is due to the neutral hydrogen in our own galaxy, and its column density ($N_{\rm{H,G}}$) is listed in Table \ref{t:summary}. The other is the intrinsic obscuration from the gas at the redshift of the source ($N_{\rm{H,in}}$). $N_{\rm{H,in}}$ is a free parameter in the spectral fitting as well as the photon index $\Gamma$ of the power law. However, due to the low signal-to-noise ratio (the number of photons detected by the PN and MOS detectors are listed in Table \ref{t:cstat}), we were not able to constrain both $N_{\rm{H,in}}$ and $\Gamma$ simultaneously in several cases. Therefore for sources with low S/N, we alternately froze $N_{\rm{H,in}}$ and $\Gamma$ and let the other parameter vary. The same methodology was applied when adding an additional thermal component into the power-law continuum. 

\subsubsection{Absorbed power-law model}
\begin{figure}
\centering
\begin{tabular}{cc}
\epsfig{file=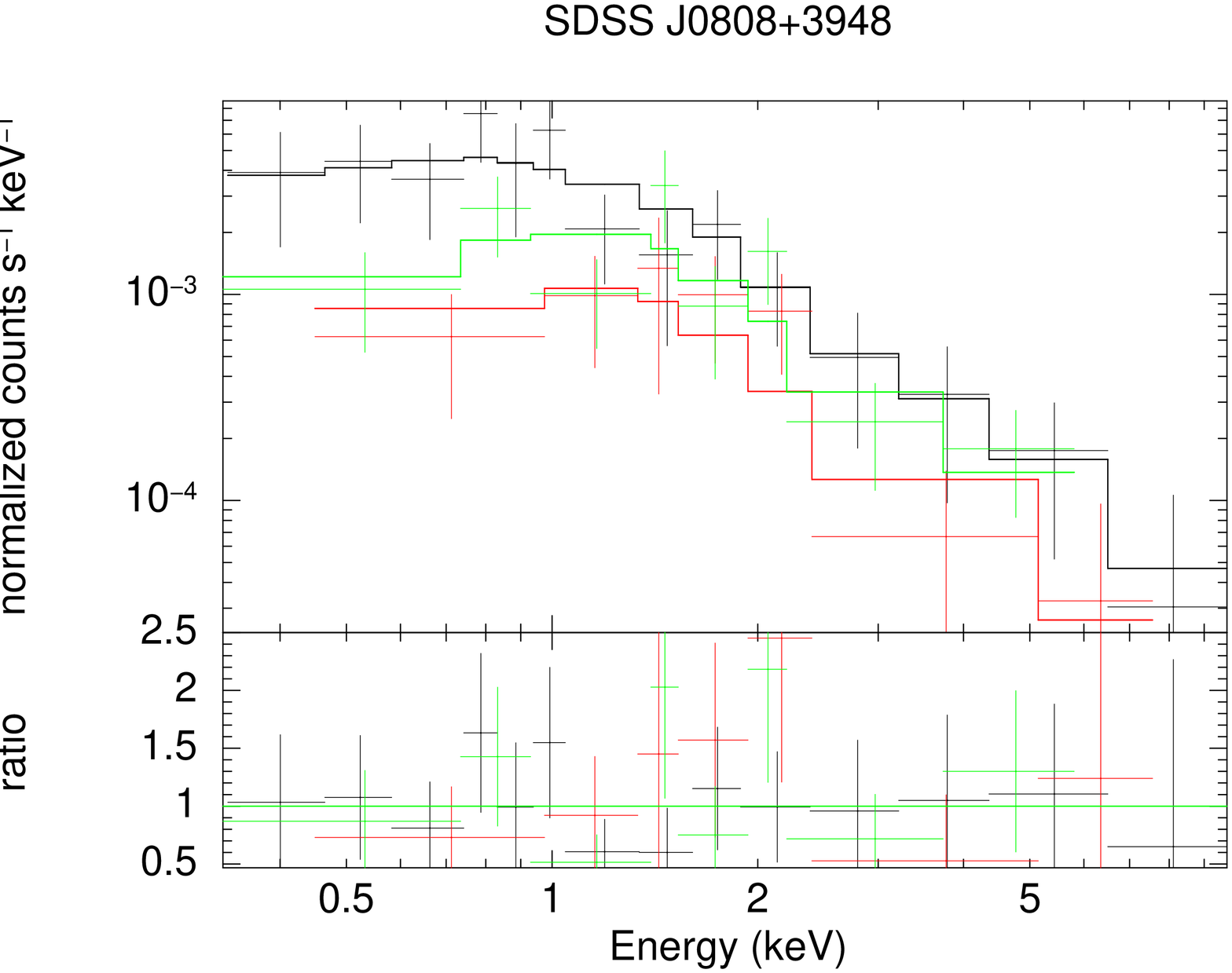,width=0.4\linewidth,angle=-90,clip=} &
\epsfig{file=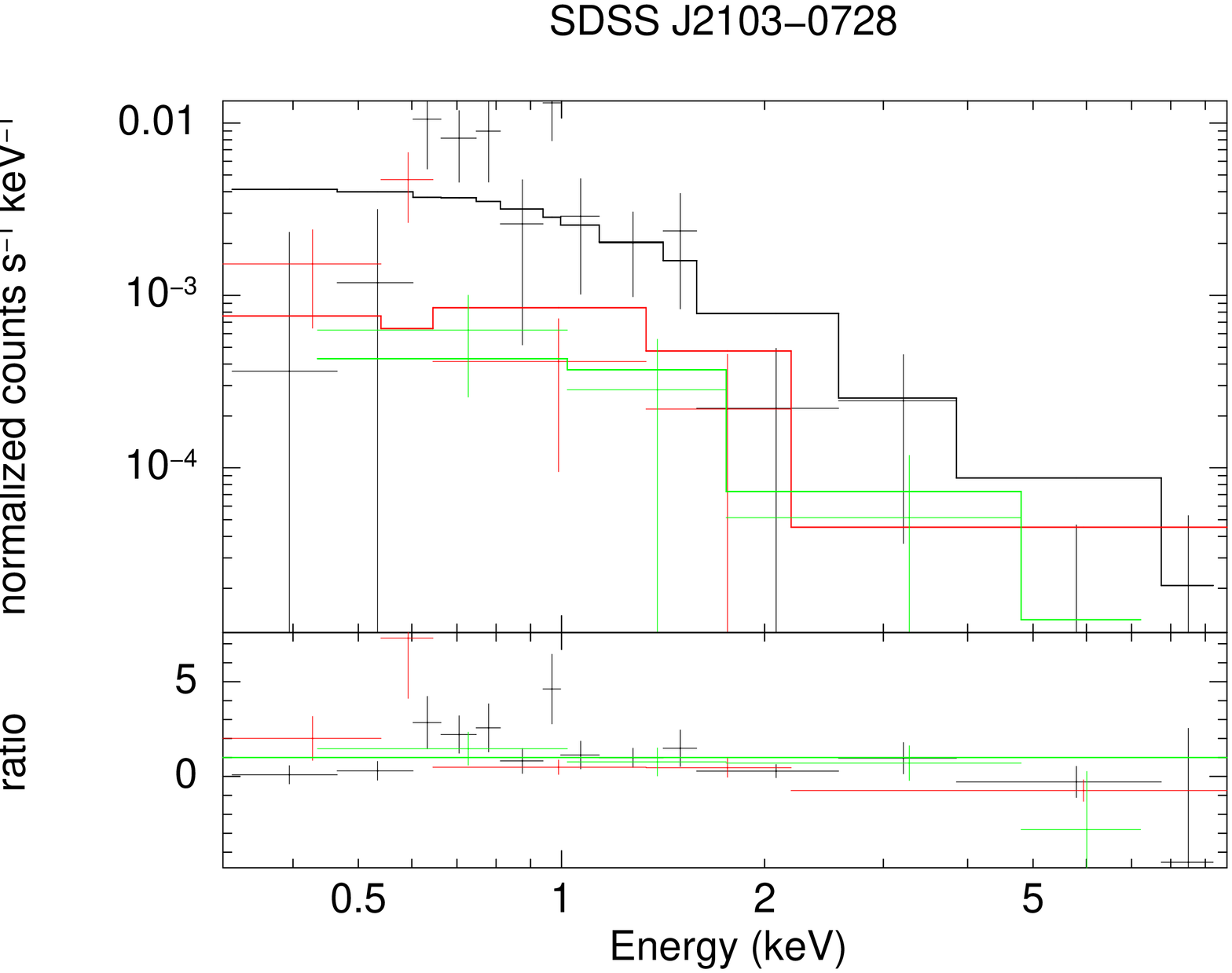,width=0.4\linewidth,angle=-90,clip=} \\
\epsfig{file=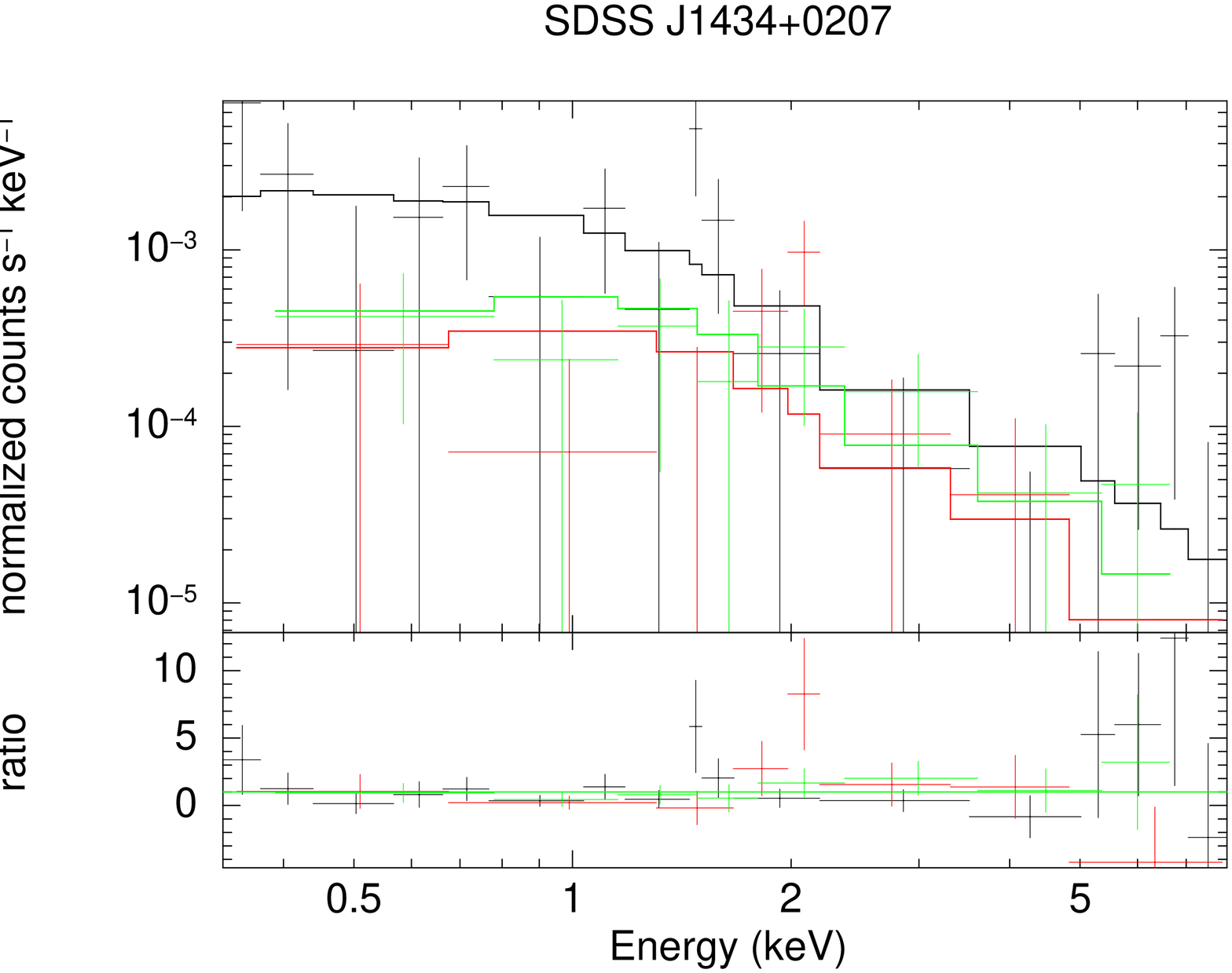,width=0.4\linewidth,angle=-90,clip=} &
\epsfig{file=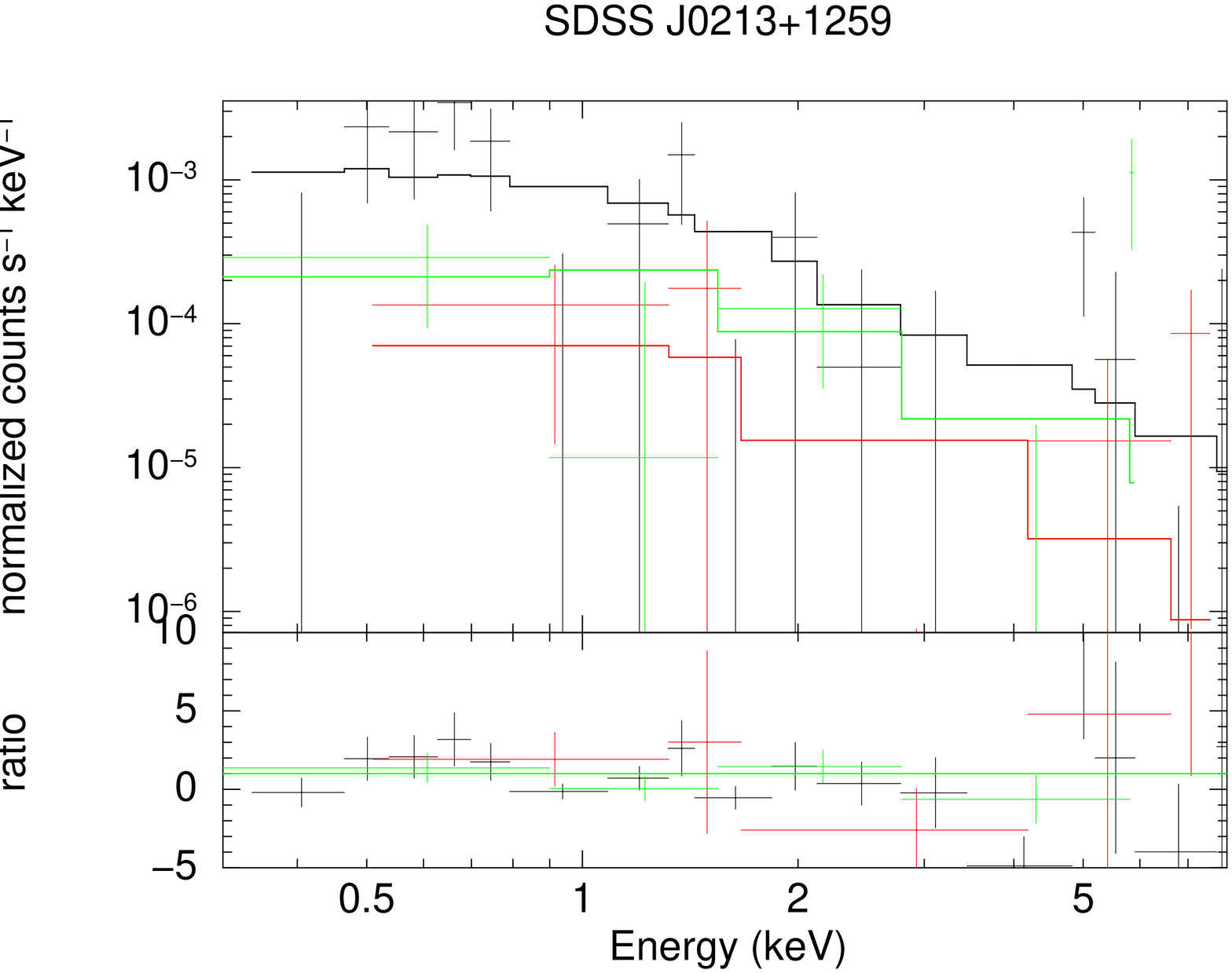,width=0.4\linewidth,angle=-90,clip=} \\
\epsfig{file=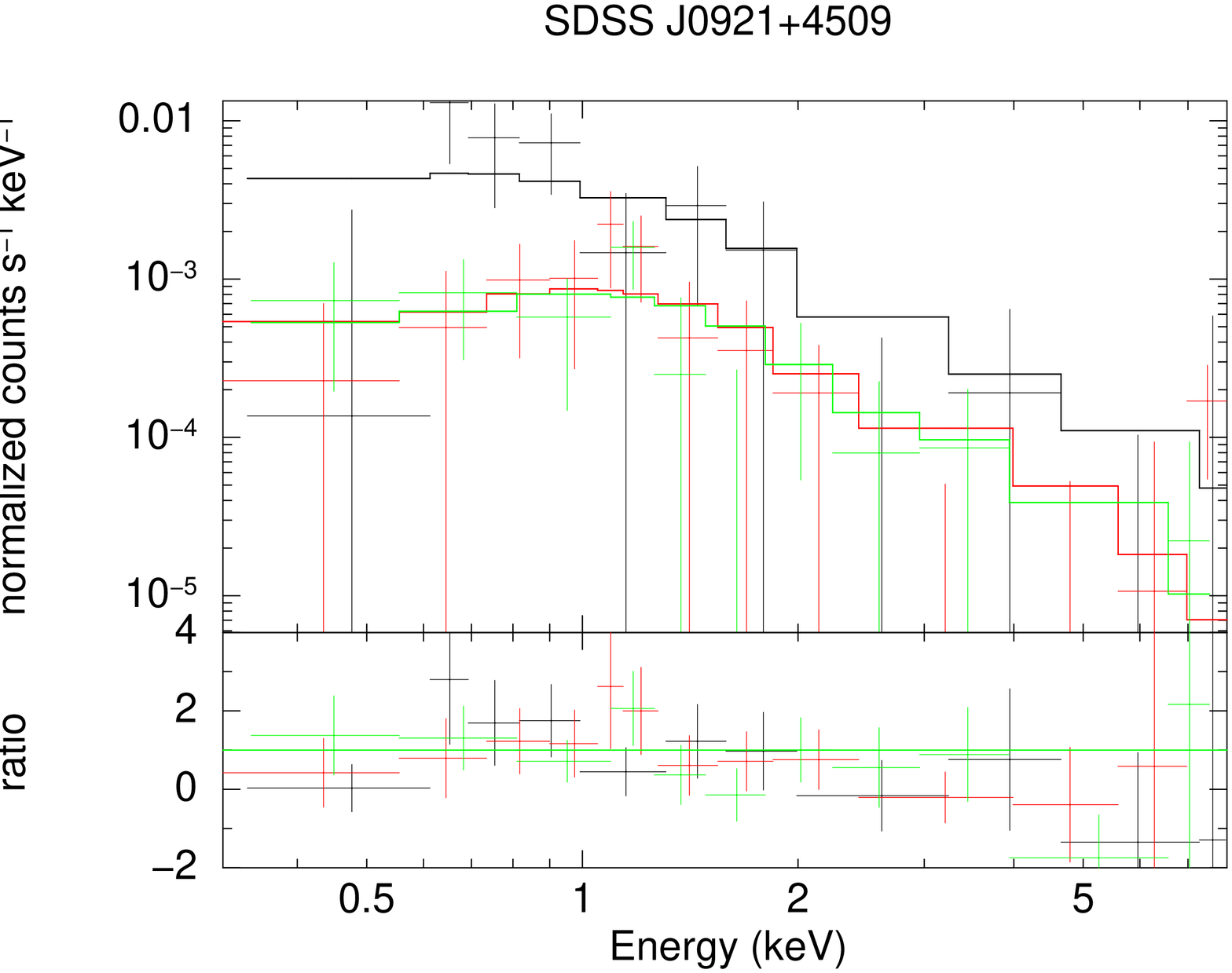,width=0.4\linewidth,angle=-90,clip=} &
\epsfig{file=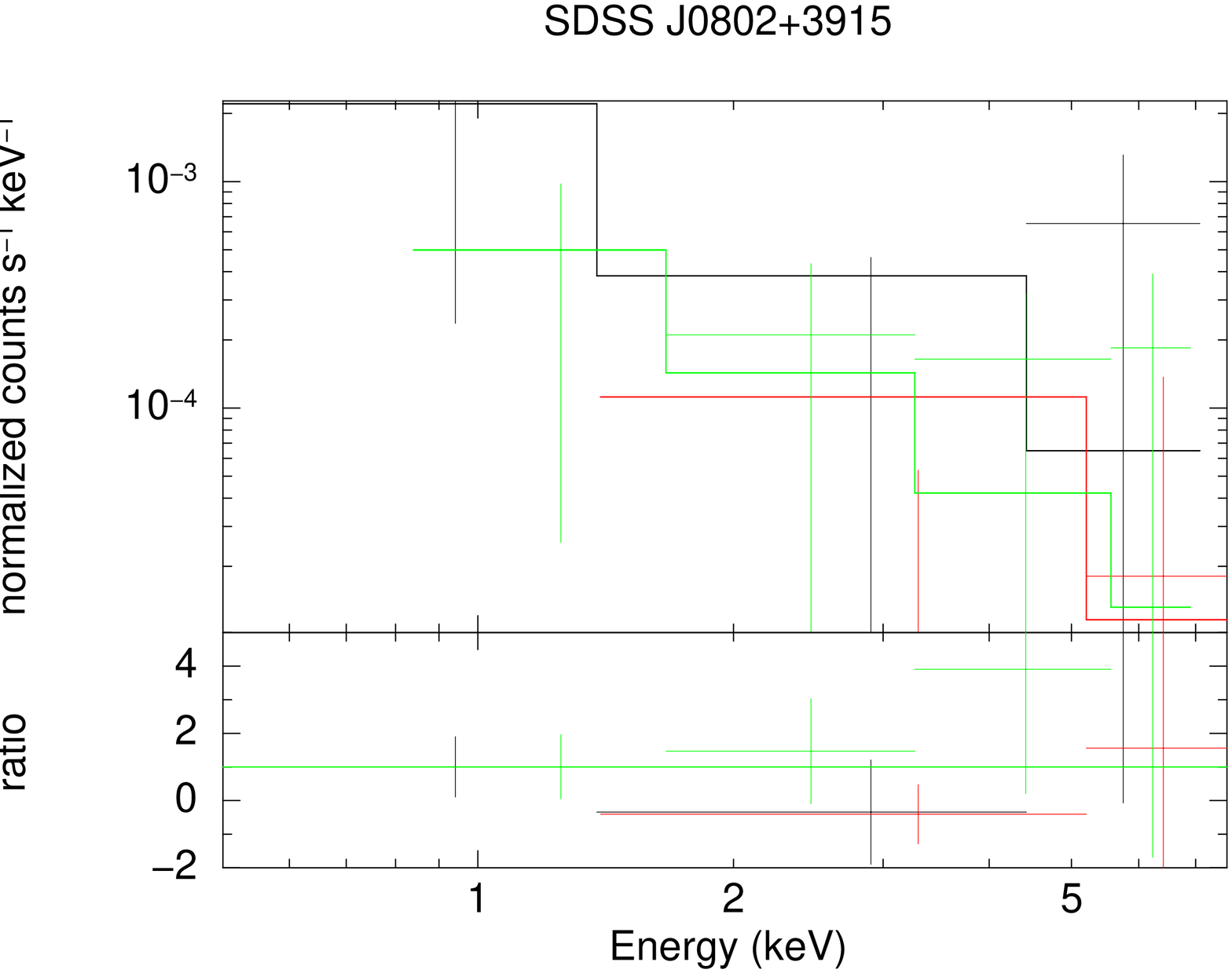,width=0.4\linewidth,angle=-90,clip=} 
\end{tabular}
\caption{X-ray spectral plots of each LBA-composite fitted with $C$-statistic, and the ratios between the model and data of the fit are shown in the bottom panels. The colors of black, red and green indicate the data in detectors of PN, MOS1 and MOS2, respectively.\label{f:cstat}}
\end{figure}

As mentioned above we fitted the spectrum in two ways to derive the power-law photon index and the intrinsic absorption: First, we freed the power-law photon index and assumed no absorption at the redshift of the source ($N_{\rm{H,in}}=0$ cm$^{-2}$) and that the continuum power-law spectra is only absorbed due to the Galactic column density (listed in Table \ref{t:summary}). Again in many cases there was not sufficient signal for the absorption and slope to both be free parameters.   We list the photon indices derived from this simple absorbed power-law model in column 3 in Table \ref{t:cstat}, as well as the corresponding $C$-statistic in column 4. Second, the spectra were fitted with the power-law model this time with the intrinsic absorption as a free parameter while the photon index was fixed as $\Gamma=1.9$, which is a mean value for X-ray detected LBGs \citep{laird06} and is also a typical spectrum for an unobscured AGN. The values of $N_{\rm{H,in}}$ are shown in column 5 in Table \ref{t:cstat}. The errors of photon indices in column 3 and the upper limits of intrinsic absorptions in column 5 were calculated at a confidence level of 90\% (i.e., single interesting parameter $\Delta C$=2.7). We also list in Table \ref{t:cstat} their derived luminosities in soft X-ray (0.5-2.0 keV) and hard X-ray (2-10 keV) bands (column 7 and 8, respectively), and the ratio of hard X-ray and [O III] luminosities (column 9). The associated uncertainties for luminosity correspond to the 90\% confidence range of normalization of the continuum power-law component in spectral fit. See Fig. \ref{f:cstat} for the spectral fits of each source with a fixed photon index $\Gamma=1.9$. 

\begin{figure}
\centering
\begin{tabular}{cc}
\epsfig{file=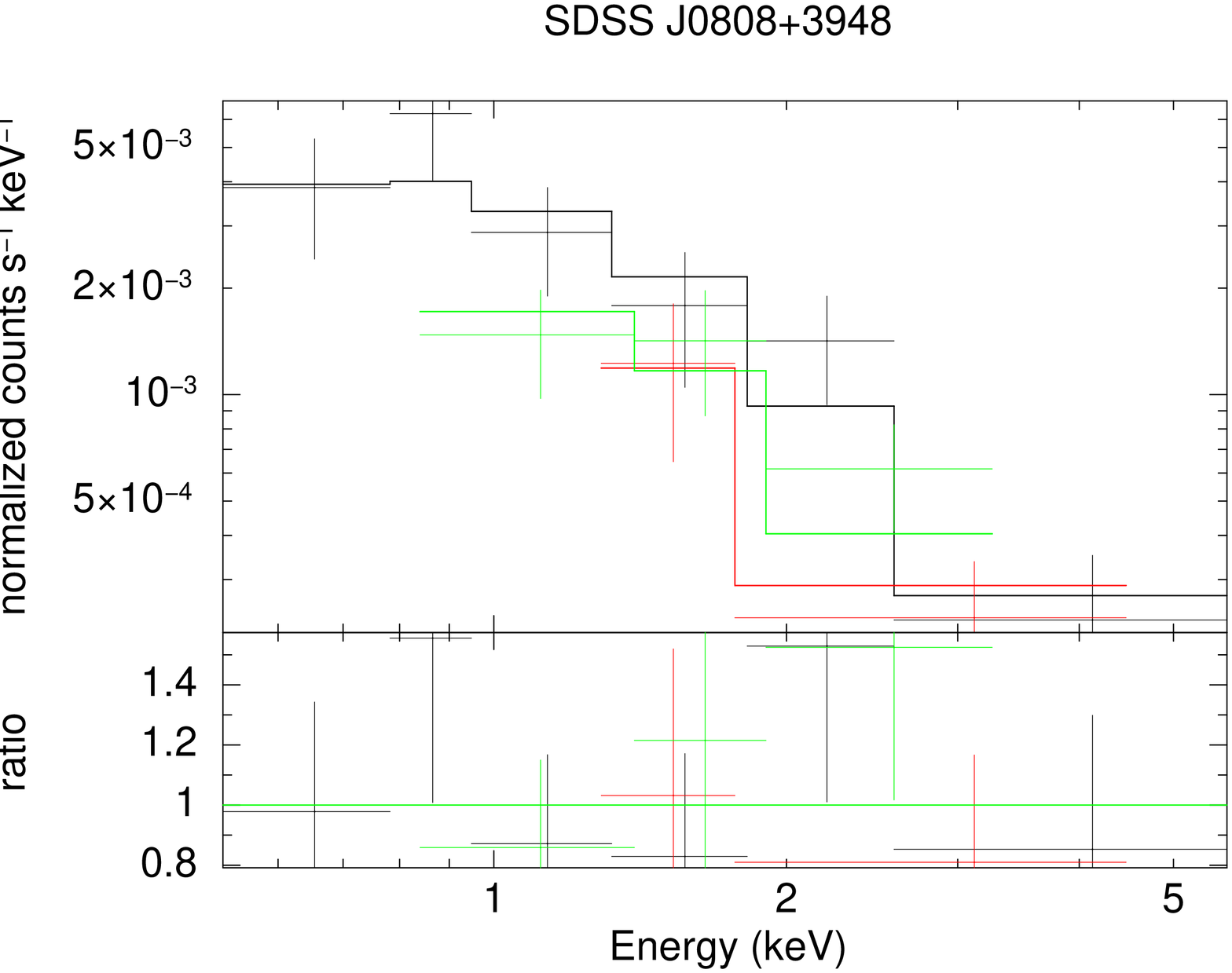,width=0.4\linewidth,angle=-90,clip=} &
\epsfig{file=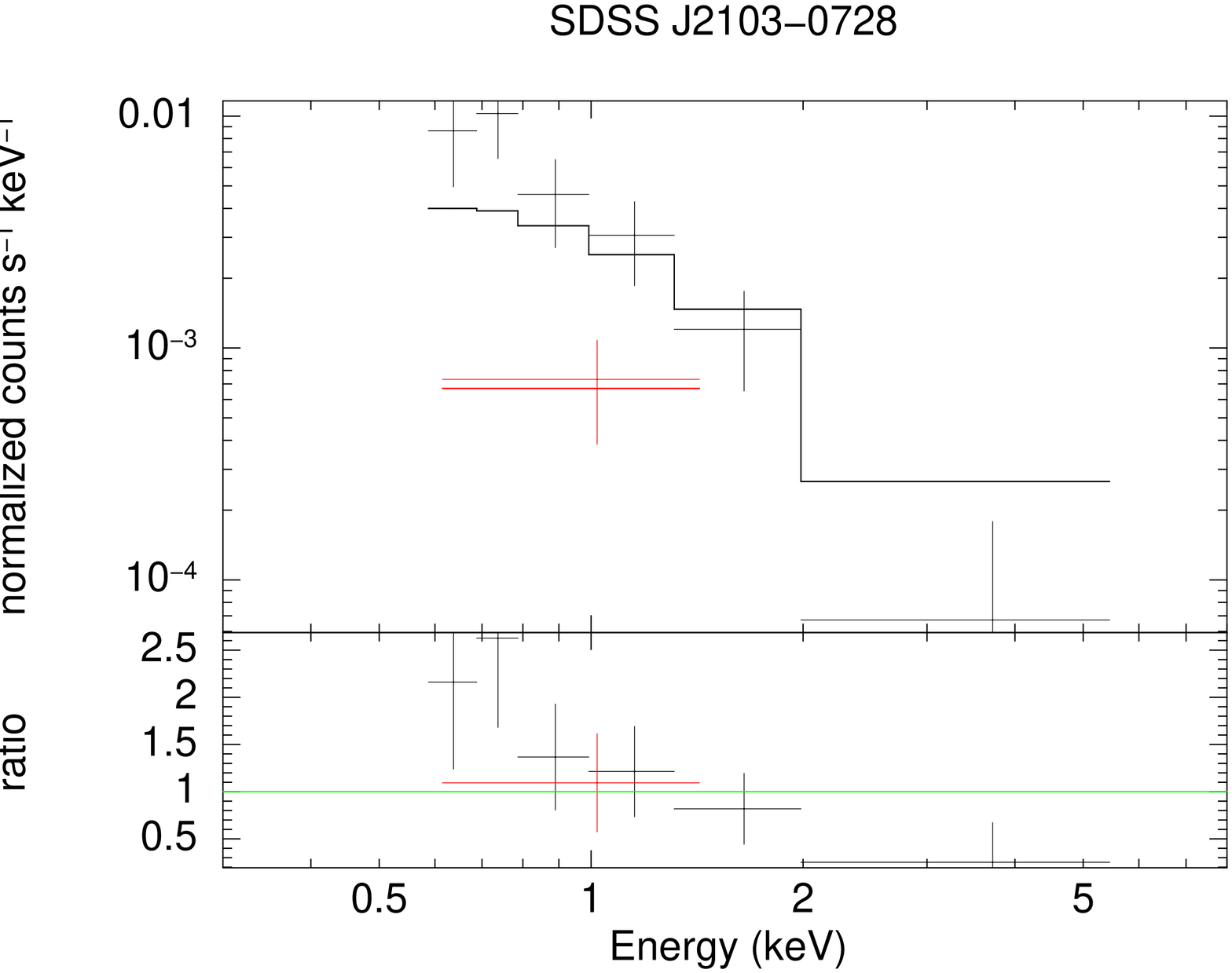,width=0.4\linewidth,angle=-90,clip=} \\
\epsfig{file=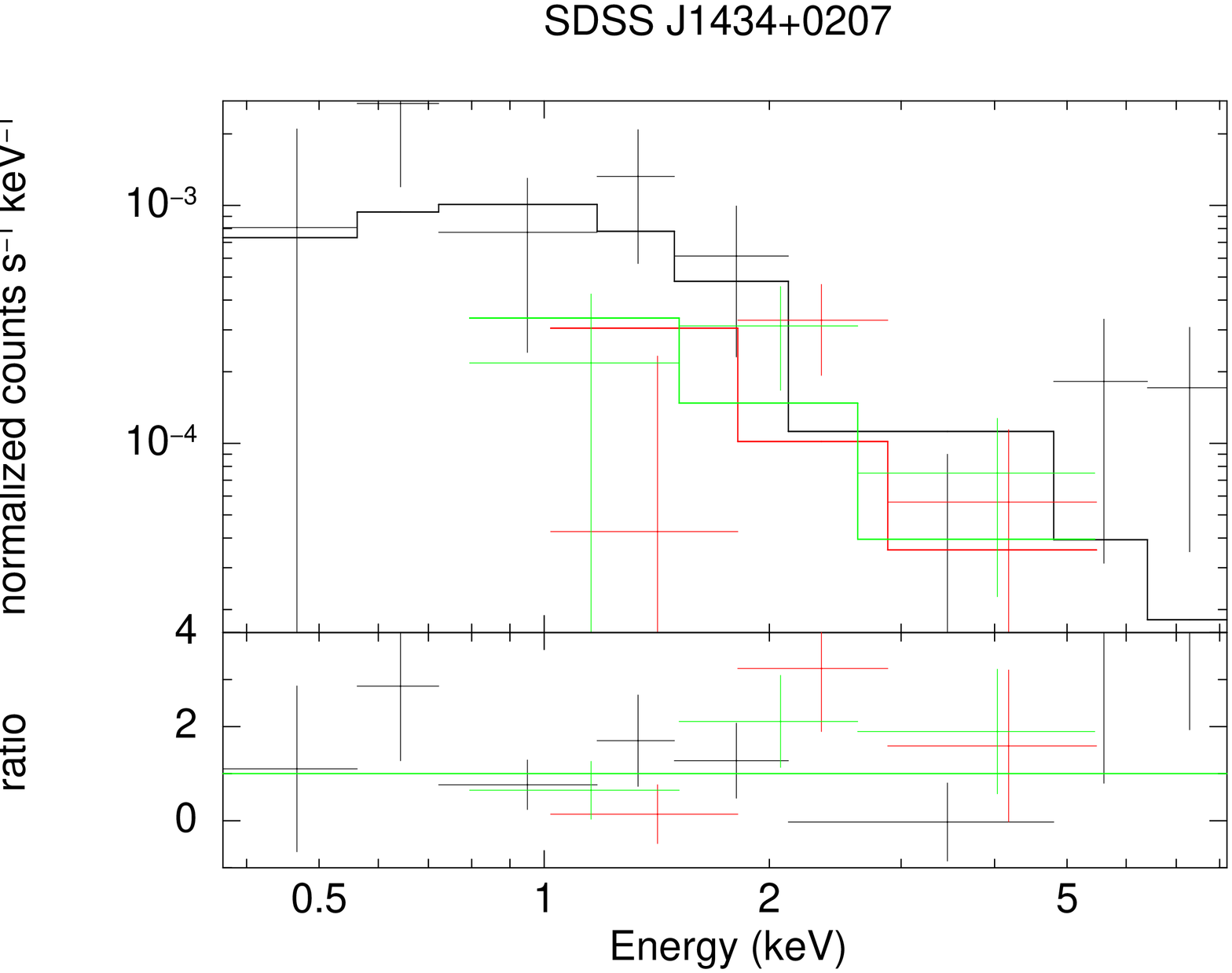,width=0.4\linewidth,angle=-90,clip=} &
\epsfig{file=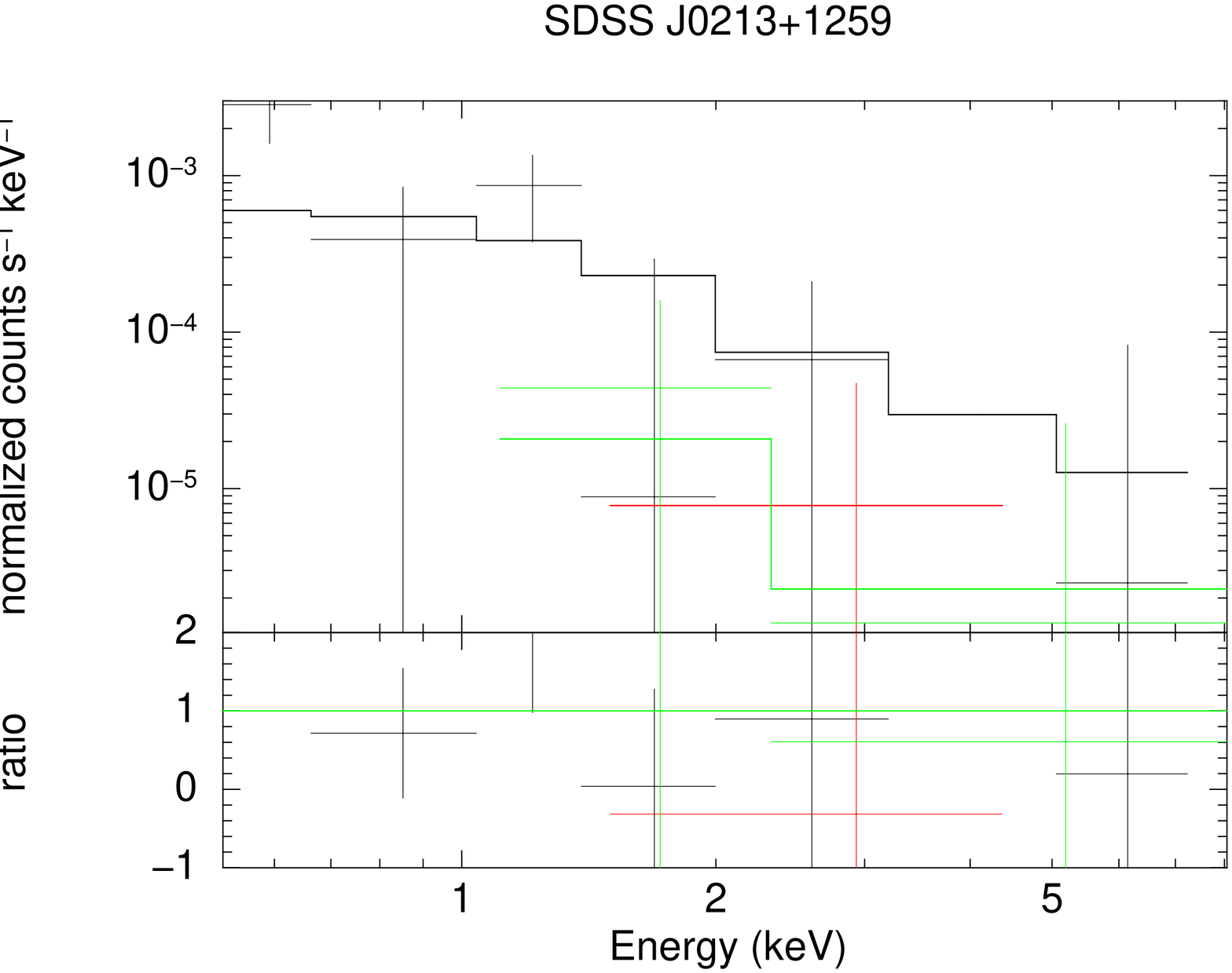,width=0.4\linewidth,angle=-90,clip=} \\
\epsfig{file=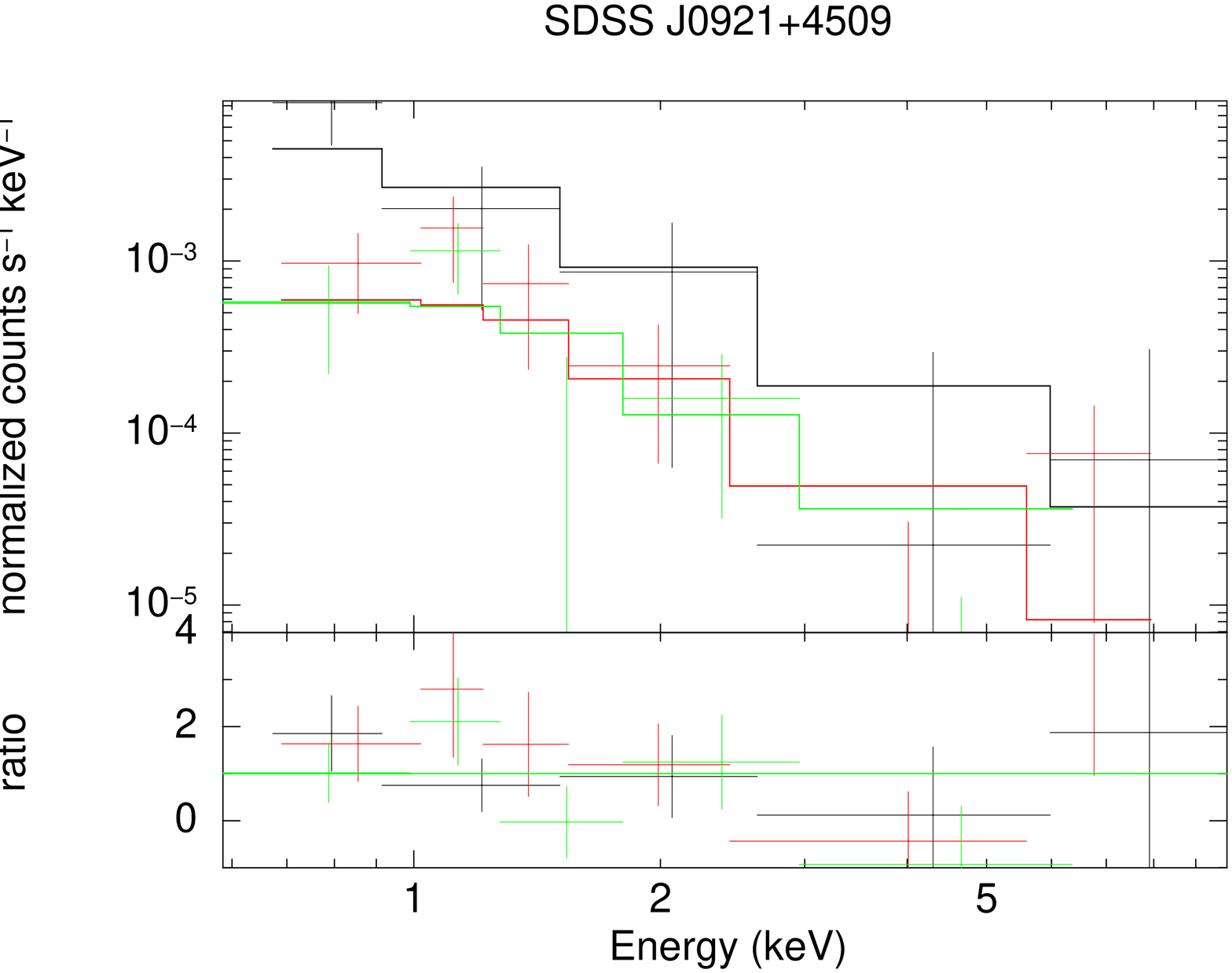,width=0.4\linewidth,angle=-90,clip=} &
\end{tabular}
\caption{X-ray spectral plots of each LBA-composite fitted with $\chi^2$-statistic, and the ratios between the model and data of the fit are shown in the bottom panels. The colors of black, red and green indicate the data in detectors of PN, MOS1 and MOS2, respectively. The plot for SDSS~J0802+3915 is not shown due to insufficient counts for $\chi^2$-statistic. \label{f:chi}}
\end{figure}

We also fitted the X-ray spectra of our targets using the $\chi^2$-statistic to see how different the fits are compared with those by $C$-statistic. The fits were performed in the same way as those using the $C$-statistic above. The spectra were grouped in 10 counts per bin for both PN and MOS detectors, and the spectra of MOS1 and MOS2 were combined in the fits. Table \ref{t:chi} lists the photon index and intrinsic column density of each source derived while the other one is frozen in the spectral fitting. SDSS~J0802+3915 is excluded due to insufficient bins for fitting after grouping into 10 counts per bin. Also, spectral plots are shown in Fig. \ref{f:chi}, which are fits with photon indices fixed at 1.9. Fitting the spectra using $\chi^2$  does not make any improvements or significant difference in estimating the absorption and photon index compared to the $C$-statistic method. Thus, we use the X-ray luminosity derived from the $C$-statistic fit in the following discussion.

Additionally, as SDSS~J0808+3948, SDSS~J1434+0207 and SDSS~J0922+4509 have relatively more photons detected, we also fitted their spectra by thawing both the parameters of photon index and intrinsic absorption (using both the $C$ and $\chi^2$ statistics). Table \ref{t:both} shows the values of these parameters and their associated errors or upper limits. Since we have two interesting parameters in the fits, the errors at the 90 per cent confidence level are calculated according to $\Delta \chi^2$ or $\Delta C$=4.6.

\subsubsection{Power-law plus thermal model\label{thermal}}

The LBAs are the sites of strong starbursts, so we would expect to see thermal X-ray emission from the hot gas that is a characteristic features of such systems \citep[e.g.,][]{grimes05,grimes06,grimes07}. We therefore fit models in which we added a thermal component to the absorbed power-law model aforementioned. We used the thermal plasma model \texttt{apec} in \texttt{XSPEC} instead of the older model,\texttt{vmekal},  used in \cite{grimes07}. The \texttt{apec} model has three free parameters: the plasma temperature, the metal abundance (in solar units) and the normalization. Due to the low number of photon counts, we fixed the abundance at the solar value. The parameter of temperature and normalization were free in fitting. However, due to insufficient photon counts, we fixed the temperature at 0.5 keV for SDSS~J0808+3948, SDSS~J0922+4509 and SDSS~J0802+3915 which is a typical value for starburst galaxies \citep[e.g.,][]{grimes05,grimes06,grimes07}. Here we kept the photon index fixed at 1.9 and allowed the intrinsic absorption at the redshift of the LBA to vary. The results are shown in Table \ref{t:apec}, which lists the temperature of plasma, the column density, and the derived luminosities of both the thermal and power-law component, as well as the $C$-stat for comparison with the single power-law fitting. The errors for parameters were calculated according to $\Delta C=4.6$, and the uncertainty of luminosity was proportional to the corresponding error of normalization of each component. For the three LBAs where the temperature was permitted to vary the temperature ranged between 0.3 and 0.6 keV. The thermal component luminosities were on the order of $10^{41}$ erg s$^{-1}$, with $10^{40}$ erg s$^{-1}$ for SDSS~J0808+3948 and about $10^{42}$ erg s$^{-1}$ for SDSS~J0802+3915. The power-law component luminosity in the hard X-ray range (2-10 keV) is within a factor of 2 of those derived from the single power-law model spectral fit mentioned above. This power-law plus thermal model fit makes no statistically significant difference in the derived column density compared to that of the single power-law model.

Since the signal-to-noise of the \xmmn~ spectra is low, we attempted to constrain any thermal component by simultaneously fitting the spectra of all LBAs together assuming that they have approximately the same hot gas temperature, i.e., we tied the temperature parameters together between the fits but let the \texttt{apec} normalization vary for each LBA. We also fixed the power-law index at 1.9, and tied the normalization of power-law component in our fit (although note that overall normalization for each LBA was allowed to be free). The temperature given by this joint fit is 0.56 keV, and its 90\% confidence level is 0.43$-$0.69  keV. We list the intrinsic column density, the thermal luminosity and the soft and hard X-ray luminosity contributed by the power-law component in Table \ref{t:apecjoint}. We see that the column densities and thermal luminosities given by this joint fit are consistent with those in the individual fits, but with improved constraints.

\begin{figure}
\centering
\begin{tabular}{cc}
\epsfig{file=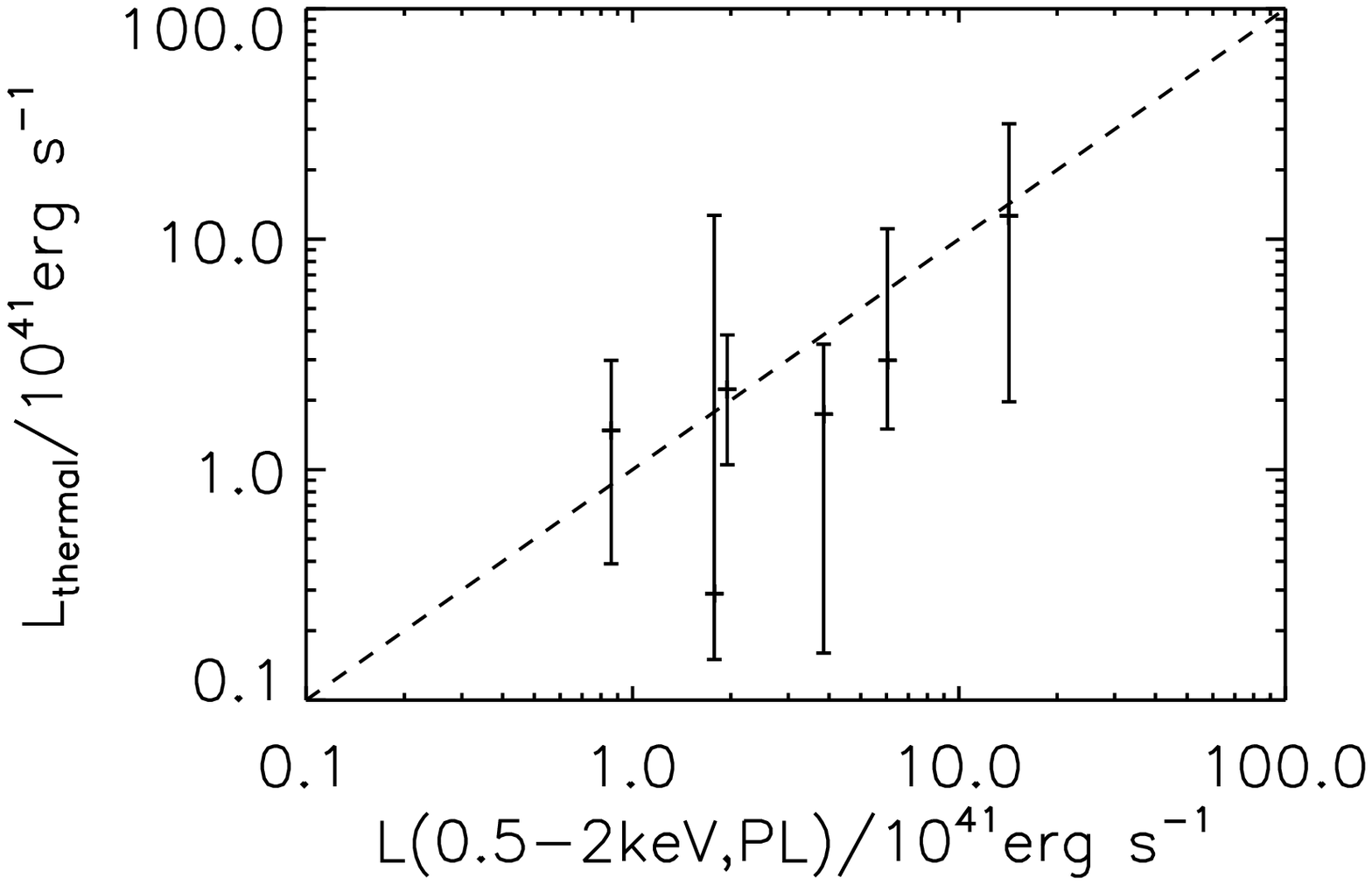,width=0.5\linewidth,angle=0,clip=} &
\epsfig{file=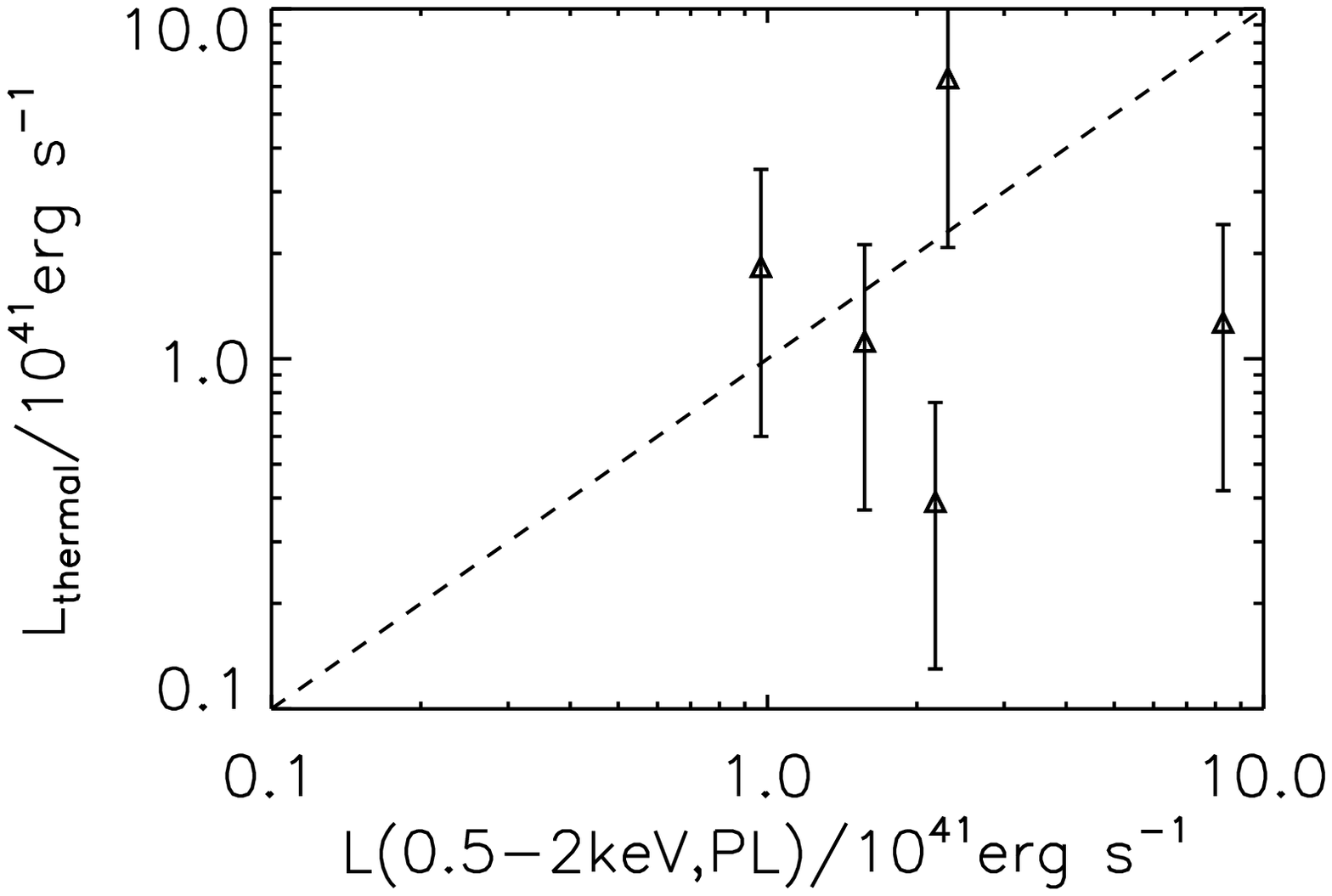,width=0.5\linewidth,angle=0,clip=} 
\end{tabular}
\caption{Comparison of contributions to the soft X-ray luminosity between power-law and thermal emission components in the power-law plus thermal spectral fits. Left panel shows the results from the individual fits (Table \ref{t:apec}), and the right panel shows those from the joint fit (Table \ref{t:apecjoint}). The error bar for soft X-ray luminosity is not shown as it makes the plot unwieldy. SDSS J0214+1259 has a negligible 0.5-2 keV power-law luminosity due to its Compton-thick obscuration estimated in the simultaneous joint fit, so it is not included in the right panel. The dashed line indicates where both values equate. \label{f:lxlapec}}
\end{figure}

In Fig. \ref{f:lxlapec}, we compared the two compositions of soft X-ray luminosity, i.e., the thermal and power-law luminosities in 0.5-2.0 keV, where the dashed line indicates where both values equate. This comparison was done separately for both the cases in which the spectra were fit individually and those that were fit simultaneously. These are shown in the left and right panel of Fig. \ref{f:lxlapec} respectively. The results derived from the individual spectra fits imply that these two mechanisms play a nearly equal role in soft X-ray emission, except in J0808+3948 where the power-law soft X-ray luminosity is about one dex higher than that of thermal emission. For the joint spectral fit, the luminosities from the two emission mechanisms are comparable, within the range of 1 dex. The obscuration of SDSS J0214+1259 is Compton-thick in the joint fit, and the power-law luminosity at 0.5-2 keV is negligible (not shown).

\subsection{Star formation rates}

We list the UV luminosity for each LBA measured from the OM observation in Table \ref{t:uv}. We derive the SFR from our UV luminosities using the formula from \cite{rosa02}, which is based on evolutionary synthesis models of starbursts by assuming a Salpeter initial mass function (IMF). For consistency with \cite{overzier09}, we divide the Rosa-Gonz\'{a}lez et al. relation by a factor of 1.5 to reflect the use of a Kroupa rather than a Salpeter IMF (as described in \cite{calzetti09}). The resulting formula is as follows:

\begin{equation}
\rm{SFR(UV)}=9.3\times 10^{-29}L_{\nu}(\rm{erg~s}^{-1}\rm{Hz}^{-1})~~\rm{M}_{\odot}/\rm{yr}.
\end{equation}

In Table \ref{t:sfr} we compare our SFR derived from the UV luminosity with that derived from a combination of the 24$\mu$m and H$\alpha$ luminosities \citep[see][for details]{overzier09}. Due to the uncertain infrared SEDs, no $K$-correction was applied to the 24$\mu$m luminosity in \cite{overzier09}, and the actual SFRs might be somewhat higher than their derived values. The UV-derived SFR is lower than that derived from the IR continuum plus optical emission lines. This is consistent with the fact that the UV photons suffer significant dust attenuation and are re-emitted in the IR \citep[see also][]{overzier11}.

We note that any AGN contribution to the UV continuum is negligible in these objects. This is demonstrated most directly by \textit{HST} COS FUV spectra of two of our targets \citep{heckman11}.

\section{Discussion\label{discussion}}

As we described in the introduction, the optical emission-line spectra of our sample of LBAs suggest that these may be composite objects consisting of an intense starburst and an obscured AGN. However, this evidence is ambiguous. \cite{overzier09} have discussed this in detail and suggest as an alternative that the optical spectra of these galaxies could reflect extreme conditions in the ionized gas related to the presence of a starburst with extreme properties.

The X-ray data we have presented provide important new insight into the nature of these objects. We begin by comparing the amount of detected hard and soft X-ray emission in the LBAs compared to pure starbursts with the same star-formation rates (as traced by the far-IR luminosity). We will then compare the LBAs to typical Type 1 and Type 2 AGN in terms of the ratios of the X-ray to [O III] and mid-IR to [O III] luminosities.

\subsection{X-ray properties compared to starbursts\label{4.1}}

The far-infrared (FIR) luminosity is considered to be a reliable indicator of the global SFR of star forming galaxies. In particular, \cite{ranalli03} found that the 2-10 keV X-ray luminosity has a tight linear correlation with the FIR luminosity by investigating a sample of local star forming galaxies. We show our LBAs in the $L_{\rm{2-10~keV}}-L_{\rm{FIR}}$ diagram (see Fig. \ref{f:sfrlx}) with the Ranalli relation, which is indicated by the solid line with two dashed lines indicating a factor of 2 above and below. The hard X-ray luminosities of the LBAs are taken from Table \ref{t:cstat}, which are derived from the absorbed power-law model fit with the photon index fixed at 1.9. For comparison, we also show the sample of Luminous Infrared Galaxies (LIRG) from \cite{iwasawa09} and \cite{lehmer10}, where Iwasawa et al's sample was divided into two groups of AGN and ``Hard X-ray Quiet" galaxies (HXQ) indicated by diamond and asterisk symbols respectively, and Lehmer et al's sample was indicated by triangle symbols\footnote{\cite{lehmer10} suggest that the X-ray emission from high-mass X-ray binaries may be suppressed by heavy extinction in some of the HXQ, LIRGs and ULIRGs}. We can see that our LBAs are about 0.5 to 1 dex above the Ranalli et al. relation, and even more displaced from the non-AGN LIRGs. This implies that a normal star forming process would only account for a minority of the hard X-ray emission of these LBAs.

We also fit a $\log (L_{\rm{2-10~keV}})-\log (L_{\rm{FIR}})$ correlation from our LBA sample by applying the survival analysis \citep[\texttt{ASURV} Rev 1.2;][]{isobe90}, which gives
\begin{equation}
\log (L_{\rm{2-10~keV}})=(0.41\pm0.35)\times \log (L_{\rm{FIR}})+(23.42\pm 11.28)
\end{equation}
The dash-dotted line shows the best fit for the correlation.

We find similar results in the soft X-ray band.
We fit the correlation between the soft X-ray (0.5-2 keV) and FIR luminosity of our LBA sample, and compared it with the result of \cite{ranalli03}. The survival analysis gave a free slope fit
\begin{equation}
\log (L_{\rm{0.5-2~keV}})=(0.42\pm0.24)\times \log (L_{\rm{FIR}})+(22.89\pm 10.76)
\end{equation}

Besides the linear slope fit, \cite{ranalli03} gave a free-slope fit of $L_{\rm{0.5-2~keV}}\propto L^{0.87\pm0.08}_{\rm{FIR}}$ for star forming galaxies. In Fig. \ref{f:softlfir}, the dash-dotted line indicates the best-fit of LBAs, and the Ranalli curve is shown as the dashed line below it. As in the case of the hard X-ray emission, only part of the soft X-ray flux is likely to be contributed by normal processes associated with star formation.

Given that the LBAs suffer from less dust extinction than typical local starbursts with similar star-formation rates \citep{overzier11}, one might worry that the high ratio of X-ray to far-IR luminosity seen in Figures 7 and 8 is due to the incomplete conversion of far-UV into far-IR light. This is not the case. As discussed in \cite{overzier11}, the star-formation rates derived from the observed far-UV are much smaller than those derived from the far-IR data, implying that the far-IR is a good proxy for the star-formation rate. Our results in section 3.2 and Figure 6 further support this.

\subsection{X-ray Properties compared to AGN}

The results from \S \ref{4.1} are consistent with interpreting the optical spectra as indicating the presence of an AGN. The best tracer of both Type 1 and Type 2 AGN in the optical spectra is the [O III]$\lambda$5007 emission-line. In this case, we would expect that the amount of hard X-ray emission from the LBAs would be consistent with the values found for AGN with similar [O III] luminosities.

In Fig. \ref{f:lxlo}, we plot a histogram of the ratio of the hard X-ray (2-10 keV) luminosity obtained from the power-law model spectral fit of our LBAs to its observed [O III] luminosity. We also show the empirical distribution of the luminosity ratios of Type 1 (dashed blue line) and Type 2 AGN (dot-dashed red line) in this plot \citep{heckman05b,lamassa10b}. The ratios of hard X-ray to [O III] luminosities in the LBAs are too low (by about an-order-of-magnitude) to be consistent with the values seen in Type 1 AGN. However, they do lie within the relatively broad range of values seen in Type 2 AGN. The relatively weak hard X-ray luminosity of Type 2 AGN is due to obscuration of the hard X-rays by the high column densities of gas associated with the AGN torus. In contrast, the [O III] line is formed in the 100 to 1000 pc-scale narrow line region and is not affected by the torus surrounding the black hole.

At first sight, the presence of significant obscuration of the hard X-ray emission in the LBAs would not be consistent with the relatively low absorbing column densities listed in Table \ref{t:cstat}. Here we see that the column density derived from the spectral fitting of an absorbed power-law gives the 90\% confidence level upper limit at the order of $10^{21}$ cm$^{-2}$ ($10^{22}$ cm$^{-2}$ for SDSS J0802+3915). These columns are much too low to attenuate the hard X-ray emission by $\sim$ an order-of-magnitude. Such a seeming paradox is often seen in Type 2 AGN and is most likely due to fitting an overly idealized X-ray spectral model. More realistic models (e.g. partial covering models and/or models with reprocessing) \citep{turner97, lamassa09,lamassa10b} can reconcile the observed X-ray spectrum with heavy obscuration.

Alternatively, the actual absorbing column densities derived from the fits to the X-ray spectra may be correct. In this case, the low ratio of hard X-ray to [O III] luminosities relative to unobscured AGN would require that the [O III] line is predominantly produced by the starburst rather than the AGN. As discussed by \cite{overzier09}, this possibility can not be excluded.

Assuming for the moment that the relatively high ratios of X-ray to far-IR luminosity indicate that an AGN is present in these LBAs, our results still do not establish whether this AGN is more luminous and heavily obscured in hard X-rays or less luminous and relatively unobscured. An additional diagnostic of an obscured AGN is a strong iron K$\alpha$ emission-line with a large equivalent width of about one keV \citep[e.g.][]{lamassa09,lamassa10a}. In Fig. \ref{f:cstat}, we show the comparison between the data and the spectral fit from an absorbed power-law. We find that there is a large data-to-model ratio at the Fe K$\alpha$ energy in 1434+0207, 0214+1259 and 0802+3915. We used a power-law plus Gaussian model to fit the PN spectra of 1434+0207 and 0214+1259 (insufficient photon counts for 0802+3915). We removed the 5-8 keV photons first to determine the power-law index, then froze this parameter and added the 5-8 keV photons with a Gaussian component. The line energy is 6.6 keV for 1434+0207 and 6.2 keV for 0214+1259, and the spectral plots are shown in Fig. \ref{f:iron}. The plots indicate the possibility of an iron emission line in these LBAs. However, the photon counts are too few to give good estimates of the equivalent width of iron lines.

In order to further investigate the presence of a Fe line in our sample, we loaded the spectra of all 6 LBAs in \texttt{XSPEC} to construct an average spectrum due to the small number of photons of each LBA (effectively stacking the spectra in model space). This was done in the same way by which we fitted the thermal component simultaneously in \S \ref{thermal}. This is not a fit to a stacked spectrum, but rather a joint fit to the set of observed spectra without shifting them to their rest-frame.The averaged spectrum was modeled as a continuum power-law in the range of 3-8 keV plus a Gaussian emission line. The  intrinsic line width ($\sigma$) in the Gaussian component was fixed at 0.01 keV, and the photon indices of the continuum power-law were set to 1.9. By tying the Gaussian normalization of all data sets, we obtain the emission line energy at 6.40 keV, with a 90\% confidence region ($\Delta C$-stat=4.6 for two model components) of 6.09-6.68 keV. The equivalent width (EW) is 1.0 keV from this joint fit, and the associated 90\% confidence upper limit of EW is about 4.4 keV, where the upper limit  is obtained by scaling the confidence region of the line intensity by the best-fit value of the EW, which is likely to be overestimated \citep{yaqoob98}. However, the data has low signal-to-noise ratio, and the emission line profile is not significant. When we removed the Gaussian component in the spectral fitting, the $C$-stat only increased by 5.5 compared to the value of 109.6/102(dof) for the model with Gaussian.

We then performed the Markov Chain Monte Carlo (MCMC) simulation to test the significance of the iron emission line from the joint fit. We generated 6 chains, and the length of each chain is $5\times 10^5$. The 90\% confidence level for the line energy from the MCMC is 6.10-6.93 keV, and the estimated upper limit of EW is 2.5 keV. The histogram distribution of the line energy and normalization parameters is shown in Fig. \ref{f:mcmc}. The significance of the Fe K$\alpha$ line is calculated as the fraction of non-zero line normalization in the chains, which is found to be 95.3\%, i.e., at the $\sim 2\sigma$ level.

\subsection{Mid-IR Constraints on a Type 2 AGN}

One other test for the presence of a Type 2 AGN in the LBAs is to compare the amount of mid-IR emission expected from a such an AGN with the observed [O III]$\lambda$5007 luminosity. \cite{lamassa10a} have shown that both of these are good proxies for the bolometric luminosity of an obscured AGN (unlike the 2-10 keV X-ray luminosity).

In Figure 12 we plot histograms of the ratio of mid-IR/[O III] luminosities for the sample of Type 2 AGN in \cite{lamassa10b} and for the LBAs in this paper. The distribution for the Type 2 AGN has a mean of 1.89 dex and a dispersion of 0.57 dex compared to 2.71 and 0.34 respectively for the LBAs. All six of the LBAs lie above the mean value for the Type 2 AGN.

This suggests that if the LBAs are indeed starburst/AGN composite systems, the starburst dominates over the AGN in the mid-IR in most cases. In a future paper we will examine this issue in more detail spectroscopically using \textit{Spitzer} IRS data. Note that AGN emission from the torus peaks in the mid-IR, while the emission from dusty starbursts peaks in the far-IR (as is also the case for the LBAs). We therefore conclude that the bolometric luminosity of the LBAs is primarily due to the starburst.

As noted above, it is possible that the [O III] emission in these LBAs is primarily produced by the starburst and that the hard X-ray source suffers little attenuation. In this case, the AGN contribution to the bolometric luminosity will be even smaller.

\subsection{Constraints on the Properties of Supermassive Black Holes}

As discussed in the introduction, these LBA-composites may offer us an unparalleled opportunity to study the processes involved in the formation of supermassive black holes in the early universe \citep[e.g.,][]{elmegreen08}. In this section we will infer the properties of the supermassive black holes in our sample if galaxies using standard methods.

We will consider two possibilities. The first is the minimal case in which the hard X-ray source is not obscured. In this case, the bolometric correction to the hard X-ray luminosity in \cite{marconi04} leads to a range of AGN bolometric luminosities from 1 to 6 $\times 10^{9}$ L$_{\odot}$. The implied black hole masses then range from $\sim 3 \times 10^{4}$ to $\sim 2 \times 10^{5} (L/L_{Edd})^{-1}$ M$_{\odot}$. If these black holes are indeed radiating near the Eddington limit, the implied masses are then in the range of the ``intermediate mass" black holes found in some nearby low-mass galaxies \citep[e.g.][]{peterson05}. In the context of the high redshift universe, similar objects could represent the seeds for the subsequent growth of more massive systems.

The second possibility is that the hard X-rays are significantly obscured in these LBAs and that the [O III]$\lambda$5007 emission-line is powered by a Type 2 AGN. Adopting the bolometric correction to the extinction-corrected [O III] luminosity from \cite{kauffmann09} then yields AGN bolometric luminosities of $\sim 10^{10}$ to $\sim 2 \times 10^{11}$ L$_{\odot}$ and black hole masses of $\sim 3 \times 10^{5}$ to $\sim 6 \times 10^{6} (L/L_{Edd })^{-1}$ M$_{\odot}$. In this case, an Eddington-limited black hole would have a mass of-order 10$^{-3}$ of the DCO stellar mass \citep{overzier09}. This is similar to the mass ratio seen in present-day galactic bulges \citep{haring04, marconi03}. However, if the [O III] line is contaminated or dominated by the starburst, the AGN bolometric luminosity and implied black hole mass would be overestimated.

\section{Conclusions\label{conclusion}}

We present in this paper \xmm\ observations of 6 Lyman Break Analogs (LBAs): rare members of a local population that strongly resemble Lyman Break Galaxies at high-redshift in stellar mass, star formation rate, metallicity, size, dust extinction, and velocity dispersion. Our sample consists of LBAs whose optical spectra are consistent with a composite system consisting of both an intense starburst and a Type 2 (obscured) AGN. However, other interpretations of the optical spectra are possible. To further elucidate the nature of these objects we combine our new X-ray data with SDSS optical spectra and \textit{Spitzer} mid- and far-IR photometry. We summarize the key results as follows:

(1) We fitted their X-ray spectra with an absorbed power-law model with the parameters typically found in AGN. The column densities found from the spectral fit are less than $10^{22}~\rm{cm}^{-2}$.

(2) We found that the ratios of X-ray to FIR luminosity of the LBAs are higher than the typical starburst values by factors of three to thirty. This implies a source of X-ray emission in addition to what is seen in typical starbursts. We suggest that the most likely candidate would be an AGN.

(3) The ratios of hard X-ray (2-10 keV) to [O III] luminosity are about an order-of-magnitude smaller than in unobscured (Type 1) AGN but consistent with the broad range of values seen in Type 2 AGN. This might suggest that the hard X-rays suffer a significant amount of attenuation (as in the case of typical Type 2 AGN). This would imply that the fitted column densities derived from the simple absorbed power-law model underestimate the true attenuation (as is often the case in Type 2 AGN). An alternative interpretation is that the hard X-rays are relatively unobscured and that the [O III] line is primarily produced by an extreme starburst.

(4) To discriminate between these possibilities (and provide a confirmation of the presence of an obscured AGN), we searched for the presence of the 6.4 keV Fe K$\alpha$ emission line in our LBA sample. However, the limited photon detections and low signal-to-noise ratio do not allow us to come to a strong conclusion. The MCMC simulation gives a $\sim 2\sigma$ significance level for the iron line and a 90\% confidence upper limit of 2.5 keV for the equivalent width.

(5) We find that the ratio of mid-IR (24$\mu$m) and [O III]$\lambda$5007 luminosities of the LBAs are higher than the values seen in Type 2 AGN (by a difference of 0.8 dex in the mean). This would imply that the bolometric luminosity of these systems is primarily due to the starburst rather than an AGN. If the [O III] emission is primarily due to a starburst, this conclusion is even stronger.

(6) Given that these objects have significantly higher ratios of X-ray to far-IR luminosity than starbursts, we conclude that the most likely interpretation is that these galaxies are indeed composites of an intense starburst and an AGN (consistent with their optical emission-line spectra). If so, the implied black hole masses are of-order $10^5$ M$_{\odot}$ for Eddington-limited accretion in the case where the hard X-rays are unobscured and roughly an order-of-magnitude larger for an obscured AGN that produces the observed [OIII] emission-line. These may be good analogs to black holes forming in the dense, stellar clumps in galaxies in the early universe.

\acknowledgments

We are very grateful to the anonymous referee for helpful comments that have improved the manuscript. We also thank Marat Gilfanov for discussion. This work is supported by NASA grant NNX08AZ0G.

{}

\clearpage
\begin{table}[ht]
\setlength{\tabcolsep}{1pt}
\centering
\caption{\xmmn~observations of Lyman Break Analog targets \label{t:summary}}
\begin{tabular}{c c c c c c c c c c}
\hline\hline
Name & Observation ID & Date & Exp. Time$^{\rm{a}}$ & ${N_{\rm{H,G}}}^{\rm{b}}$ & $z$ & Mode & Filter\\
(SDSS~J...) & & (DD/MM/YY) & (ks) & & & & &\\
\hline
080844.26+394852.3 & 0553790101 & 03/10/08 & 8.7/11.5/11.5 & 5.05 & 0.091 & FF$^{\rm{c}}$ & medium \\
210358.74$-$072802.4 & 0553790201 & 18/05/08 & 8.7/11.5/11.5 & 5.41 & 0.137 & FF & thin \\
143417.15+020742.3 & 0553790301 & 20/07/08 & 13.7/22.1/22.1 & 2.80 & 0.180 & FF & thin \\
021348.53+125951.4 & 0553790401 & 08/01/08 & 19.4/23.6/23.7 & 7.06 & 0.219 & FF & thin \\
092159.38+450912.3 & 0553790501 & 27/04/09 & 3.7/20.6/22.4 & 1.55 & 0.235 & FF & medium \\
080232.34+391552.6 & 0553790701 & 24/03/09 & 1.3/6.4/6.3 & 5.15 & 0.267 & FF & thin\\
\hline
\tablenotetext{\rm{a}}{The exposure time is the good time interval after flare filtering. We list the exposure time for all three detectors (PN/MOS1/MOS2).}
\tablenotetext{\rm{b}}{Galactic column densities from the \texttt{COLDEN} tool,in units of $10^{20}$ cm$^{-2}$.}
\tablenotetext{\rm{c}}{Full frame imaging window mode.}
\end{tabular}
\end{table}

\begin{deluxetable}{cccccc}
\tablewidth{0pt}
\tablecaption{Ultra-Violet observations of Lyman Break Analog targets. \label{t:uv}}
\tablehead{ID & Filter & Corrected Rate & Magnitude & UV flux & L$_{\rm{UVW1}}$ \\
& & s$^{-1}$ & & ($10^{-16} \ \rm erg\ s^{-1}\ cm^{-2}\ \AA^{-1}$) & 10$^{43}$erg s$^{-1}$}
\startdata
SDSS~J0808+3948 & UVW1 & $2.220\pm 0.092$ & $16.34\pm 0.045$ & $10.57\pm 0.44$ & 2.70\\
SDSS~J2103$-$0728 & UVW1 & $0.938\pm 0.038$ & $17.28\pm 0.045$ & $4.45\pm 0.18$ & 2.74\\
SDSS~J1434+0207 & UVW1 & $0.475\pm 0.039$ & $18.01\pm 0.089$ & $2.26\pm 0.19$ & 2.09 \\
                & UVM2 & $0.141\pm 0.024$ & $17.90\pm 0.188$ & $3.10\pm 0.53$ &\\
SDSS~J0214+1259 & UVW1 & $0.502\pm 0.054$ & $17.95\pm 0.116$ & $2.39\pm 0.25$ & 3.85 \\
		& UVM2 & $0.099\pm 0.027$ & $18.28\pm 0.291$ & $2.18\pm 0.59$ &\\
SDSS~J0922+4509 & UVW1 & $0.882\pm 0.045$ & $17.34\pm 0.056$ & $4.20\pm 0.21$ & 7.93\\
		& UVM2 & $0.265\pm 0.025$ & $17.21\pm 0.104$ & $5.83\pm 0.55$ & \\
		& UVW2 & $0.119\pm 0.020$ & $17.18\pm 0.178$ & $6.79\pm 1.14$ &\\
SDSS~J0802+3915 & U & $0.587\pm 0.033$ & $18.84\pm 0.061$ & $1.14\pm 0.06$ &\\
		& UVW1 & $0.252\pm 0.033$ & $18.84\pm 0.061$ & $1.20\pm 0.16$ & 3.03 \\
		& UVM2 & $0.114\pm 0.019$ & $18.13\pm 0.183$ & $2.51\pm 0.42$ &\\
		& UVW2 & $0.056\pm 0.018$ & $17.99\pm 0.338$ & $3.20\pm 1.03$ &\\ 
\enddata
\tablecomments{The filters with decreasing wavelength are U (344nm), UVW1 (291nm), UVM2 (231nm) and UVW2 (212nm), and the corresponding count rates/flux conversion factors are based in white dwarf standard stars.}
\end{deluxetable}

\begin{table}[ht]
\setlength{\tabcolsep}{1pt}
\centering
\caption{[O III] and IR luminosities \label{t:loiii}}
\begin{tabular}{c c c c c}
\hline\hline
 ID & $L_{\rm{[O III],obs}}$ & $L_{\rm{[O III],corr}}$ & $L_{24\mu m}$ & $L_{\rm FIR}$ \\
& ($10^{41}$) & ($10^{41}$) & ($10^{44}$) & ($10^{44}$)\\
\hline
SDSS~J0808+3948~~ & $0.43\pm 0.01$ & 0.66 & 0.76 & $1.1\pm 0.2$ \\
SDSS~J2103$-$0728~~ & $5.00\pm 0.06$ & 12.03 & 4.8 & $8.8\pm 1.3$ \\
SDSS~J1434+0207~~ & $1.32\pm 0.03$ & 2.84 & 0.48 & $2.0\pm 0.6$ \\
SDSS~J0214+1259~~ & $0.57\pm 0.04$ & 1.22 & 1.9 & $9.1\pm 1.7$ \\
SDSS~J0922+4509~~ & $2.05\pm 0.06$ & 4.75 & 2.4 & $14.3\pm 2.2$ \\
SDSS~J0802+3915~~ & $1.54\pm 0.05$ & 4.47 & 1.2 & $6.5\pm 2.0$ \\ 
\hline
\tablecomments{[O III] luminosities are obtained from MPA/JHU catalog of SDSS DR7. The mid- and far-IR luminosities are based on \textit{Spitzer} photometry \citep{overzier11}. All luminosities are in units of erg s$^{-1}$.}
\end{tabular}
\end{table}

\clearpage
\begin{landscape}
\begin{deluxetable}{ccccccccc}
\tablewidth{0pt}
\tabletypesize{\scriptsize}
\tabletypesize{\footnotesize}
\setlength{\tabcolsep}{1pt}
\tablecaption{X-ray spectral fits using the $C$-statistic. The errors and upper limit correspond to the 90 per cent confidence level.\label{t:cstat}}
\tablehead{ID & Counts & $\Gamma$ & $C$-stat/dof & $~N_{\rm{H,in}}$
  ($\Gamma=1.9$) & $~C$-stat/dof & $L_{\rm{0.5-2
      keV}}$\tablenotemark{*} & $L_{\rm{2-10 keV}}$\tablenotemark{*} &
  $L_{\rm{2-10 keV}}/L_{\rm{[O III],corr}}$ \\
& PN/M1/M2 & & & ($10^{22}$cm$^{-2}$) & & ($10^{41} \ \rm erg\ s^{-1}$)& ($10^{41} \ \rm erg\ s^{-1}$) & \\
(1) & (2) & (3) & (4) & (5) & (6) & (7) & (8) & (9)}
\startdata
SDSS~J0808+3948 & 75/34/44 & 1.66 (1.41-1.92) & 116.11/138 & $<0.14$ & 115.82/138 & 2.11 (1.55-2.80) & 4.01 (2.94-5.33) & 6.08 \\
SDSS~J2103$-$0728 & 76/26/17 & 2.48 (2.08-2.94) & 112.78/100 & $<0.04$ & 118.60/100 & 2.93 (2.14-3.95) & 4.83 (3.53-6.51) &  0.40\\
SDSS~J1434+0207 & 86/41/43 & 1.46 (0.88-2.04) & 134.41/156 & $<0.27$ & 135.80/156 & 2.65 (1.53-4.28) & 4.41 (2.55-7.13) &  1.55\\
SDSS~J0214+1259 & 83/25/28 & 2.72 (1.91-3.68) & 133.53/122 & $<0.09$ & 136.31/122 & 2.39 (1.48-3.68) & 4.30 (2.66-6.62) & 3.51 \\
SDSS~J0922+4509 & 54/70/62 & 1.89 (1.50-2.30) & 172.57/165 & $<0.23$ & 171.02/165 & 12.35 (6.64-19.94) & 22.70 (12.21-36.66) & 4.78 \\
SDSS~J0802+3915 & 15/8/19 & 1.71 (-0.58-4.17) & 28.56/38 & $<1.37$ & 28.62/38 & 9.29 (0.81-31.96) & 16.80 (1.47-57.81) & 3.76 \\
\enddata
\tablecomments{Column (1): LBA names; column (2): numbers of counts detected by PN, MOS1 and MOS2; column (3): derived photon indices ($\Gamma$), no intrinsic absorption assumed; column (4) and (6): reduced $C$-statistic; column (5): derived intrinsic column densities at the redshifts of the sources; by freezing $\Gamma$ at 1.9; column (7): luminosities in soft X-ray band (0.5-2.0 keV); column (8): luminosities in hard X-ray band (2.0-10.0 keV); column (9): ratio of hard X-ray and extinction-corrected [O III] luminosities.}
\tablenotetext{*}{The errors of luminosities correspond to the errors of normalizations of power-law components in the spectral fit.}
\end{deluxetable}
\end{landscape}

\begin{deluxetable}{ccccc}
\tablewidth{0pt}
\tablecaption{X-ray spectral fits using the $\chi^2$-statistic. The errors and upper limit correspond to the 90 per cent confidence level.\label{t:chi}}
\tablehead{ID & $\Gamma$ & $\chi^2_{\nu}$ & $N_{\rm{H,in}}$ (for $\Gamma=1.9$) &  $\chi^2_{\nu}$\\
& & & ($10^{22}$cm$^{-2}$) & \\
(1) & (2) & (3) & (4) & (5)}
\startdata
SDSS~J0808+3948 & 1.71 (1.33-2.13) & 1.01 & $<0.24$ & 0.553 \\
SDSS~J2103$-$0728 & 3.19 (2.40-4.28) & 0.19 & $<0.08$ & 2.15 \\
SDSS~J1434+0207 & 1.24 (0.40-2.06) & 1.09 & 0.16 (0-2.44) & 1.17 \\
SDSS~J0214+1259 & $>2.05$ & 0.79 & $<0.68$ & 1.22 \\
SDSS~J0922+4509 & 2.83 (2.01-4.18) & 0.72 & $<0.18$ & 0.99 \\
\enddata
\tablecomments{Column (1): LBA name; column (2): derived photon indices ($\Gamma$), no intrinsic absorption assumed; column (3) and (5): reduced $\chi^2$; column (4): derived intrinsic column densities at the redshifts of the sources; by freezing $\Gamma$ at 1.9.}
\end{deluxetable}

\begin{deluxetable}{ccccc}
\tablewidth{0pt}
\tablecaption{X-ray spectral fits with both photon indices and column densities free parameters. The errors and upper limit correspond to the 90 per cent confidence level.\label{t:both}}
\tablehead{ID & statistic & $\Gamma$ & $N_{\rm{H,in}}$ &  $\chi^2_{\nu}$ or \\
& method & & ($10^{22}$cm$^{-2}$) & $C$-stat/dof}
\startdata
SDSS~J0808+3948 & $\chi^2$ & 1.85 (1.22-3.05) & $<0.53$ & 0.69 \\
& $C$ & 1.79 (1.35-2.50) & $<0.22$ & 0.86 \\
SDSS~J1434+0207 & $\chi^2$ & 1.23 (0.02-3.28) & $<21.8$ & 1.18\\
& $C$ & 1.41 (0.713-2.47) & $<0.35$ & 0.87 \\
SDSS~J0922+4509 &  $\chi^2$ & 4.56 (1.87-10.0) & 0.59 (0-2.91) & 0.71\\
& $C$ & 3.35 (1.94-6.56) & 0.42 (0.05-1.41) & 1.01 \\
\enddata
\tablecomments{The errors of photon indices and column densities correspond to $\Delta \chi^2$ or $\Delta C$=4.6, since there are two interesting parameters.}
\end{deluxetable}

\begin{deluxetable}{ccccccc}
\tablewidth{0pt}
\tabletypesize{\scriptsize}
\tablecaption{apec plus power-law model\label{t:apec}}
\tablehead{ID & Temperature & $N_{\rm{H}}$ & ${L_{\rm{thermal}}}^{1}$ & ${L_{\rm{0.5-2keV,PL}}}^{2}$ & ${L_{\rm{2-10keV}}}^{3}$ & $C$-stat/dof\\
& (eV) & ($10^{22}$cm$^{-2}$) & ($10^{41} \ \rm erg\ s^{-1}$) & ($10^{41} \ \rm erg\ s^{-1}$) & ($10^{41} \ \rm erg\ s^{-1}$)}
\startdata
SDSS~J0808+3948   & 500 (fixed)   & 0.10 (0.01-0.41) & 0.29 (0.15-12.69)  & 1.78 (1.09-2.74)  & 3.76 (2.31-5.79)    & 114.8/133\\
SDSS~J2103$-$0728 & 422 (276-755) & 0.17 (0-1.03)     & 1.74 (0.16-3.50) & 1.45 (0.56-2.63)  & 3.85 (1.49-6.75)    & 101.9/96\\
SDSS~J1434+0207   & 339 (217-689) & 1.07 (0.18-2.55) & 1.48 (0.39-2.98) & 0.86 (0.19-1.75)  & 6.61 (1.47-13.44)   & 129.3/152\\
SDSS~J0214+1259   & 570 (326-844) & 266.5 (13.9-)      & 2.23 (1.05-3.84) & 1.95 (1.0-4.73) & 6.20 (3.10-15.03)      & 125.7/118 \\
SDSS~J0922+4509   & 500 (fixed)   & 0.16 (0-0.65)     & 2.98 (1.50-11.10)  & 6.04 (1.13-12.51)  & 21.84 (4.10-45.23)  & 169.4/162 \\
SDSS~J0802+3915   & 500 (fixed)   & 1.96 (0.81-50.35)   & 12.64 (1.97-31.67) & 14.23 (0.23-166.63) & 40.58 (0.66-475.17) & 24.42/35 \\
\enddata
\tablenotetext{1}{Thermal emission luminosity.}
\tablenotetext{2}{Power-law luminosity in 0.5-2 keV band.}
\tablenotetext{3}{Total luminosity in 2-10 keV.}
\end{deluxetable}

\begin{deluxetable}{ccccc}
\tablewidth{0pt}
\tablecaption{apec plus power-law model in a joint fit\label{t:apecjoint}}
\tablehead{ID & $N_{\rm{H}}$ & ${L_{\rm{thermal}}}^{1}$ & ${L_{\rm{0.5-2keV,PL}}}^{2}$ & ${L_{\rm{2-10keV}}}^{3}$\\
& ($10^{22}$cm$^{-2}$) & ($10^{41} \ \rm erg\ s^{-1}$) & ($10^{41} \ \rm erg\ s^{-1}$) & ($10^{41} \ \rm erg\ s^{-1}$)}
\startdata
SDSS~J0808+3948 & 0.14 (0.07-0.28) & 0.39 (0.13-0.75) & 2.18 (1.33-3.36) & 4.87 (2.98-7.50) \\
SDSS~J2103$-$0728 & 0.35 (0.05-1.17) & 1.83 (0.60-3.48) & 0.97 (0.37-1.76) & 3.31 (1.28-6.00) \\
SDSS~J1434+0207 & 0.59 (0.06-1.86) & 1.12 (0.37-2.12) & 1.57 (0.35-3.19) & 6.91 (1.53-14.06) \\
SDSS~J0214+1259 & 769 (45-1240) & 1.94 (0.64-3.69) & ... & 9.11 (4.67-22.1) \\
SDSS~J0922+4509 & 0.19 (0.05-0.52) & 1.27 (0.42-2.42) & 8.29 (1.55-17.17) & 19.43 (3.64-40.24)\\
SDSS~J0802+3915 & 2.94 (0.50-9.30) & 6.34 (2.08-12.05) & 2.31 (0.04-27.05) & 63.49 (1.03-743.45) \\
\enddata

\tablenotetext{1}{Thermal emission luminosity.}
\tablenotetext{2}{Power-law luminosity in 0.5-2 keV band.}
\tablenotetext{3}{Total luminosity in 2-10 keV.}
\end{deluxetable}

\begin{deluxetable}{ccc}
\tablewidth{0pt}
\tablecaption{SFRs derived from UV luminosities. \label{t:sfr}}
\tablehead{ID & SFR(UVW1) & SFR(H$_{\alpha}+24\mu m$)\tablenotemark{*}  \\
& M$_{\odot}$yr$^{-1}$ & M$_{\odot}$yr$^{-1}$}
\startdata
SDSS~J0808+3948 & 6.5 & 16.1 \\
SDSS~J2103$-$0728 & 6.7 & 108.3 \\
SDSS~J1434+0207 & 5.1 & 20.0 \\
SDSS~J0214+1259 & 9.3 & 35.1 \\
SDSS~J0922+4509 & 26.1 & 55.1 \\
SDSS~J0802+3915 & 7.3 & 30.4 \\
\enddata
\tablenotetext{*}{The values of SFRs derived from combination of the uncorrected H$_{\alpha}$ luminosity and 24$\mu m$ luminosity are obtained from \cite{overzier09}.}
\end{deluxetable}

\newpage

\begin{figure}
\centering
\epsfig{file=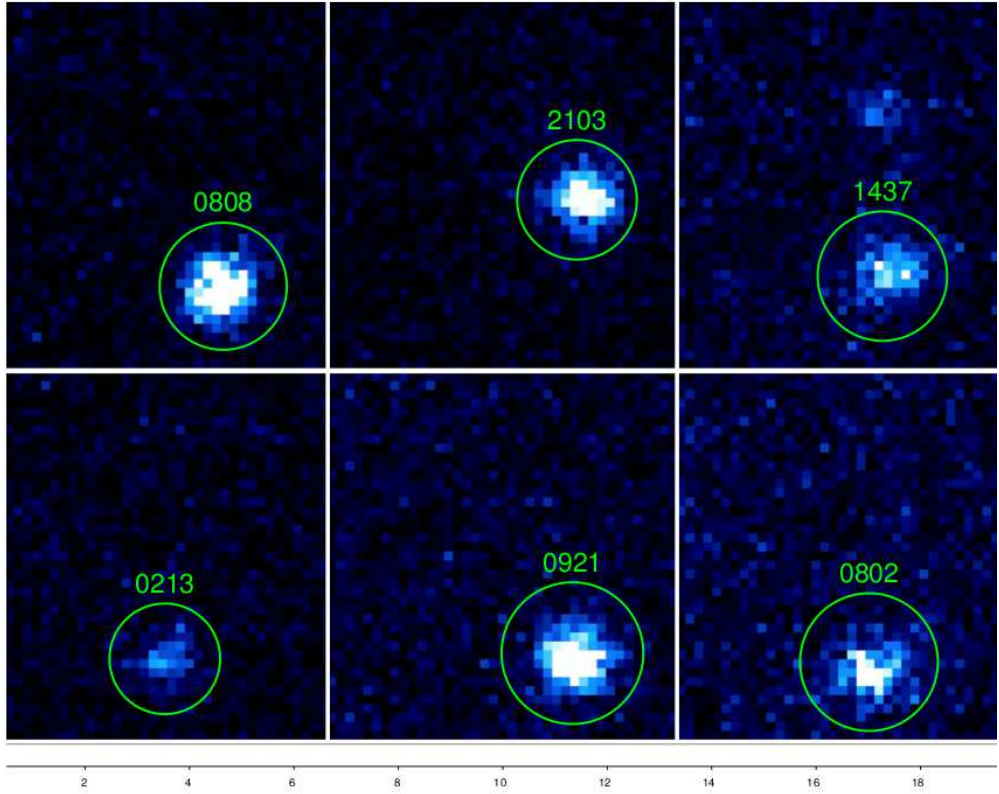,width=0.8\linewidth}
\caption{UV images (UVW1 filter) of each LBA-composite target. The size of each image is 0.6'$\times$0.6'. \label{f:uv}}
\end{figure}

\begin{figure}
\centering
\epsfig{file=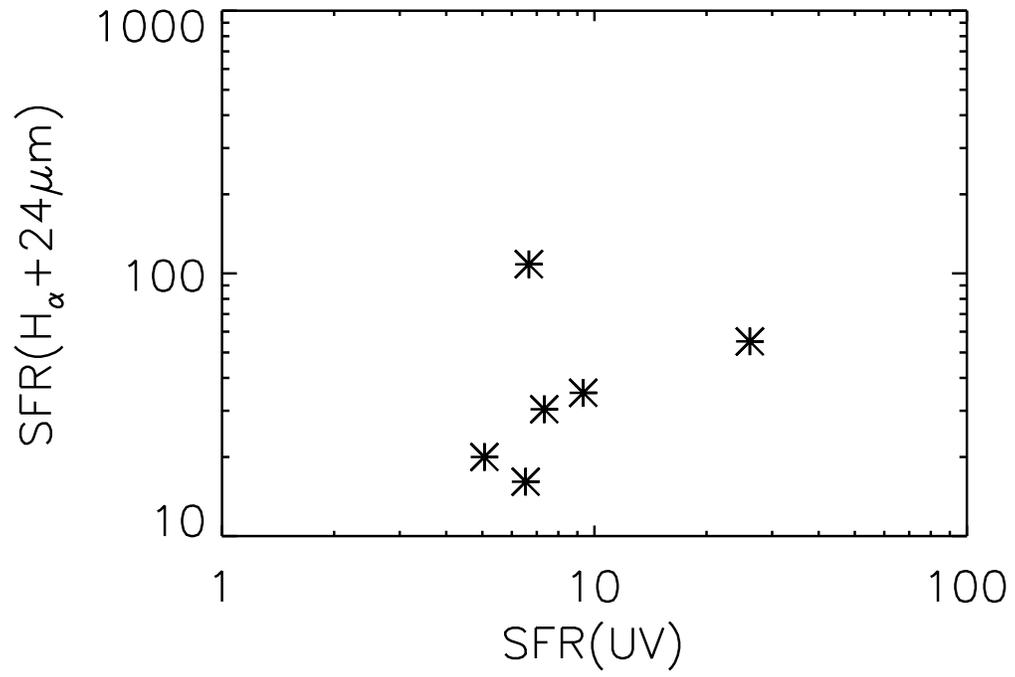,width=0.8\linewidth}
\caption{Star formion rates derived from the UV luminosity and from H$_{\alpha}+24\mu$m emission lines. The units are M$_{\odot}$ per year. \label{f:sfruv}}
\end{figure}

\begin{figure}
\centering
\epsfig{file=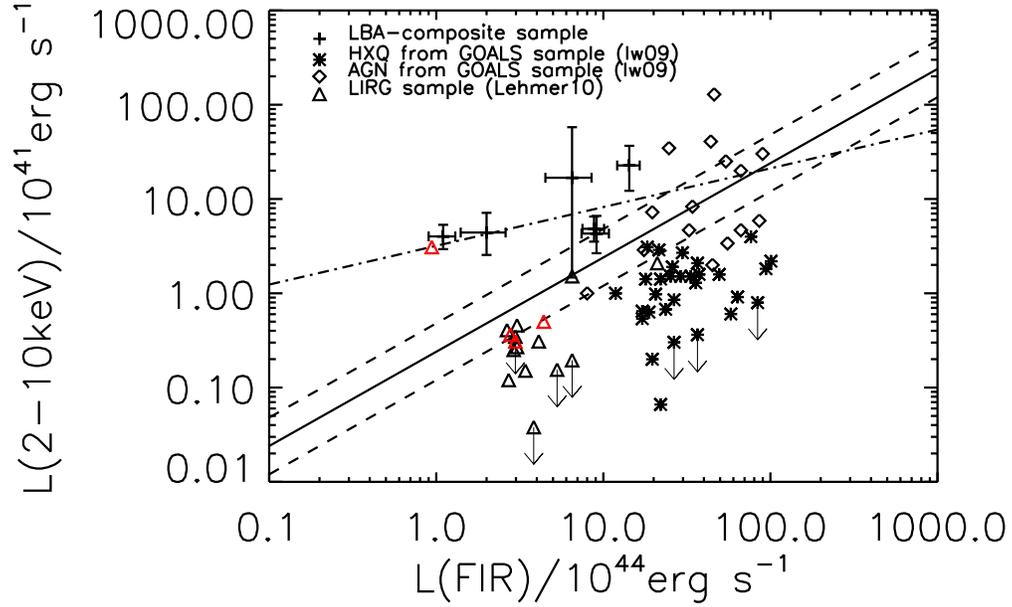,width=0.8\linewidth}
\caption{Hard X-ray luminosity versus far infrared luminosity. Our sample of LBA-composites is shown as plus signs with error bars. Others are from \protect\cite{iwasawa09} (Iw09) and \protect\cite{lehmer10} (Lehmer10) for comparison, where the diamond symbols indicate the Iw09 AGN samples, and asterisks and upper limit arrows are ``Hard X-ray Quiet" galaxies (HXQ) from Iw09, and LIRG samples from Lehmer10 are marked as triangles (triangles in red are identified as AGN). The solid line indicates the empirical hard X-ray and far infrared luminosity correlation given by \protect\cite{ranalli03}, and the dashed lines show an associated uncertainty of a factor of 2. The dash-dotted line is the correlation derived from our LBA sample by using the survival analysis.\label{f:sfrlx}}
\end{figure}

\begin{figure}
\centering
\epsfig{file=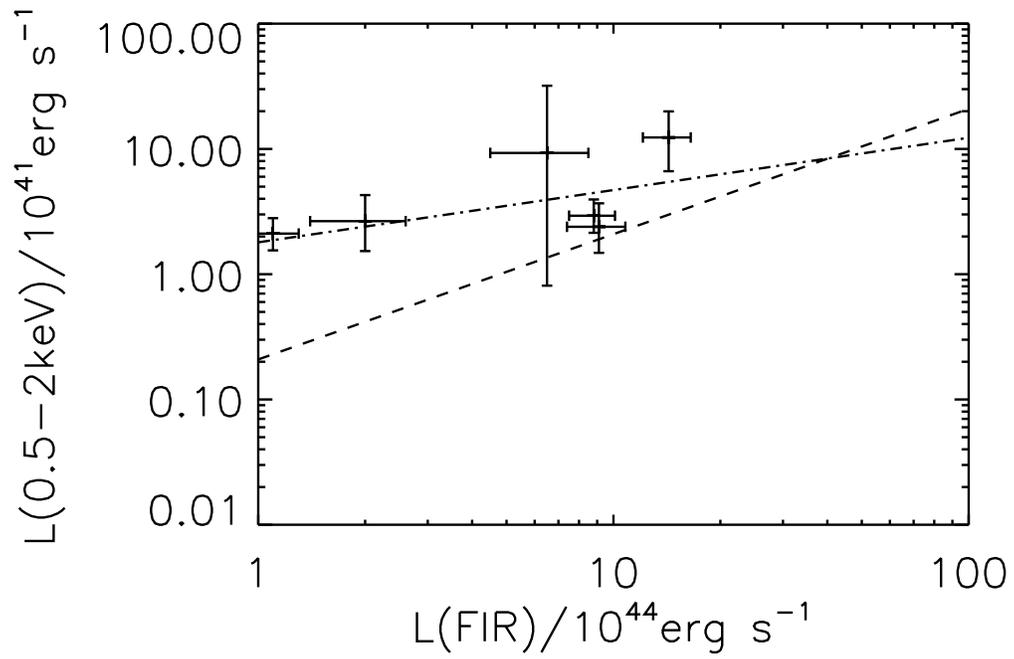,width=0.8\linewidth}
\caption{Soft X-ray (0.5-2.0 keV) luminosity versus far infrared luminosity. The dashed line is the $L_{\rm{0.5-2.0 keV}}-L_{\rm{FIR}}$ relation from \protect\cite{ranalli03}, while our best-fit correlation is indicated by the dash-dotted line. \label{f:softlfir}}
\end{figure}

\begin{figure}
\centering
\epsfig{file=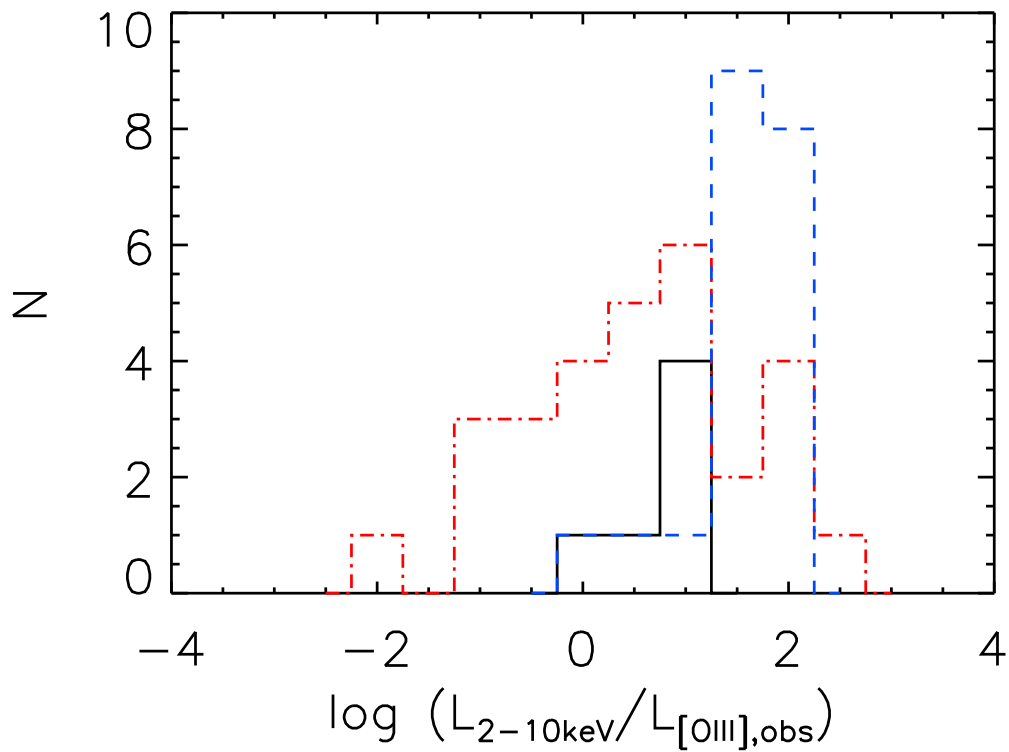,width=0.8\linewidth}
\caption{Histogram of the log of $L_{\textrm{2-10keV}}/L_{\textrm{[OIII],obs}}$ values. Our sample is represented by the solid black line. The dashed blue line and dot-dashed red line show the typical Type 1 and 2 AGNs from \cite{heckman05b}. \label{f:lxlo}}
\end{figure}

\begin{figure}
\centering
\begin{tabular}{cc}
\epsfig{file=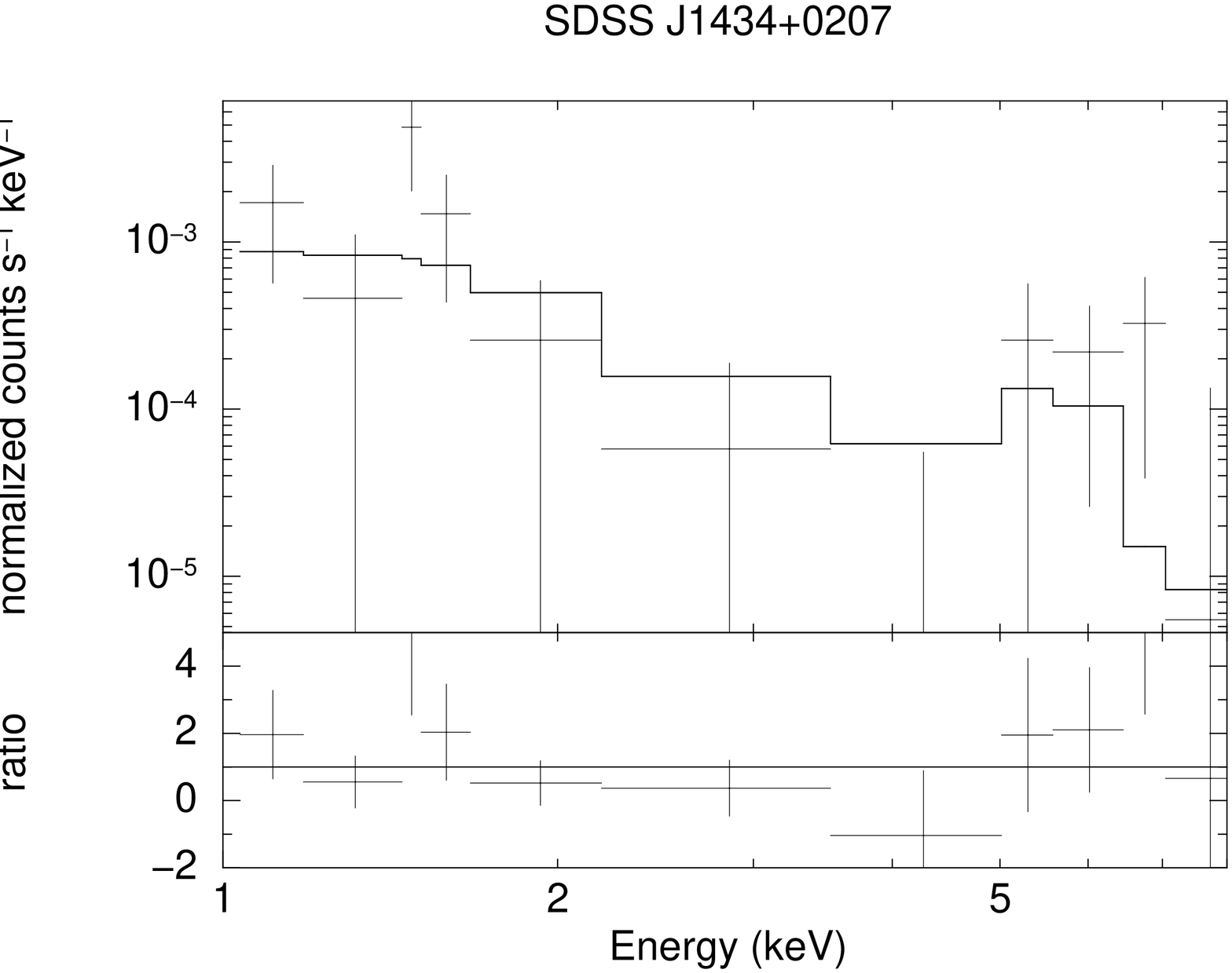,width=0.4\linewidth,angle=-90,clip=} &
\epsfig{file=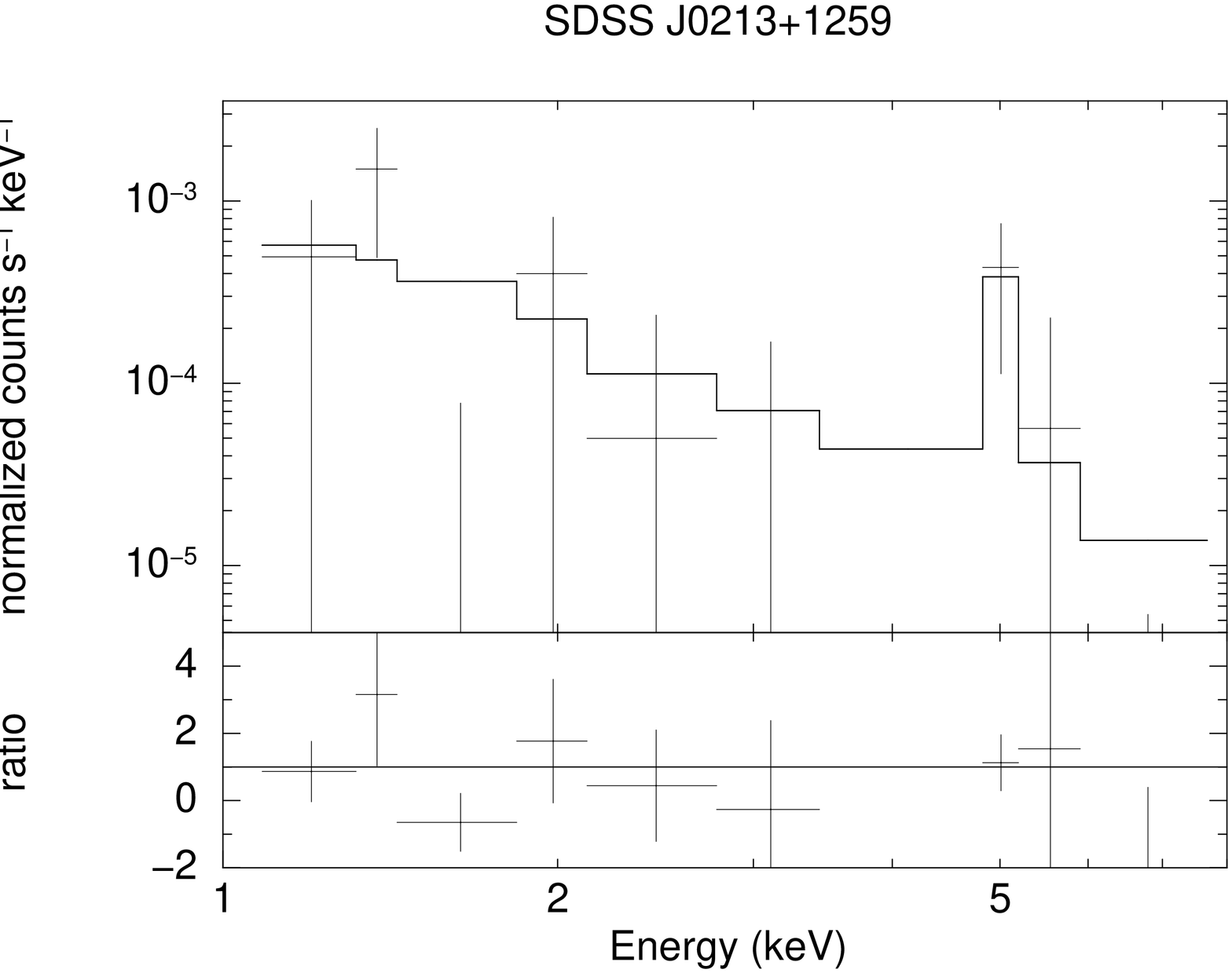,width=0.4\linewidth,angle=-90,clip=} 
\end{tabular}
\caption{Evidence for iron emission lines on power-law continuum spectra for SDSS~J1434+0207 and SDSS~J0214+1259. \label{f:iron}}
\end{figure}

\begin{figure}
\centering
\begin{tabular}{cc}
\epsfig{file=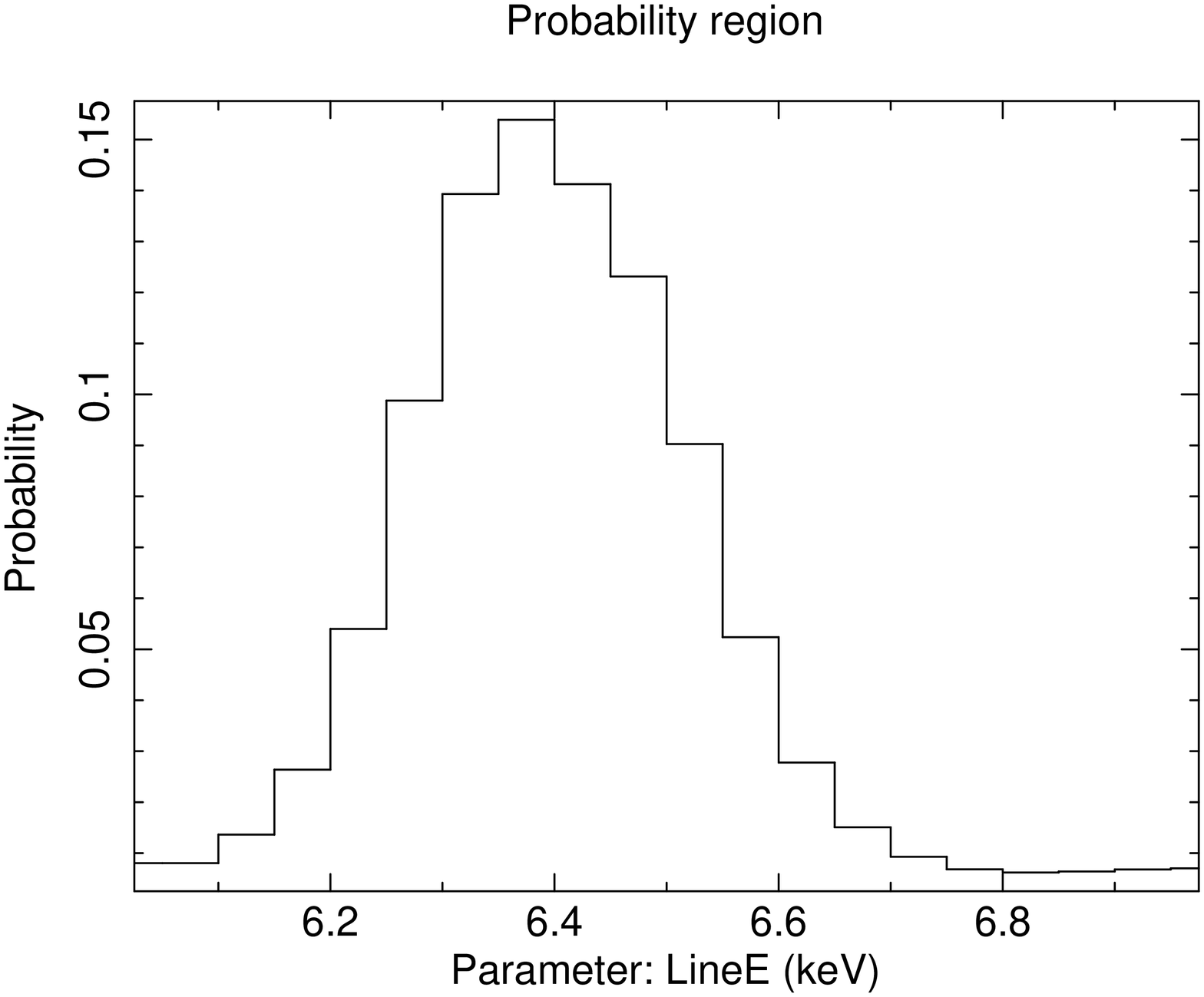,width=0.4\linewidth,angle=-90,clip=} &
\epsfig{file=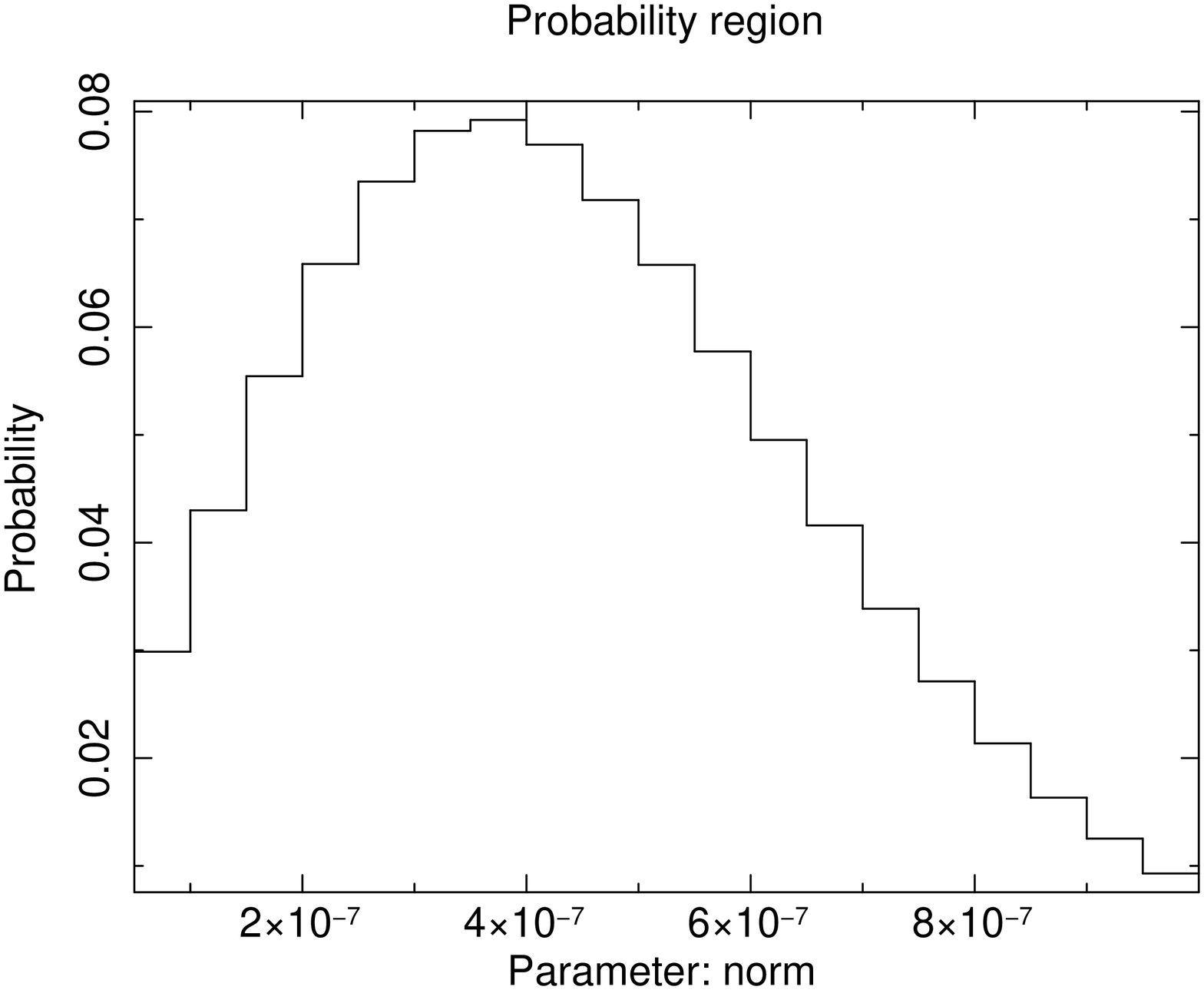,width=0.4\linewidth,angle=-90,clip=} 
\end{tabular}
\caption{Histogram distribution of the line energy and normalization parameters from the MCMC simulation. \label{f:mcmc}}
\end{figure}

\begin{figure}
\centering
\epsfig{file=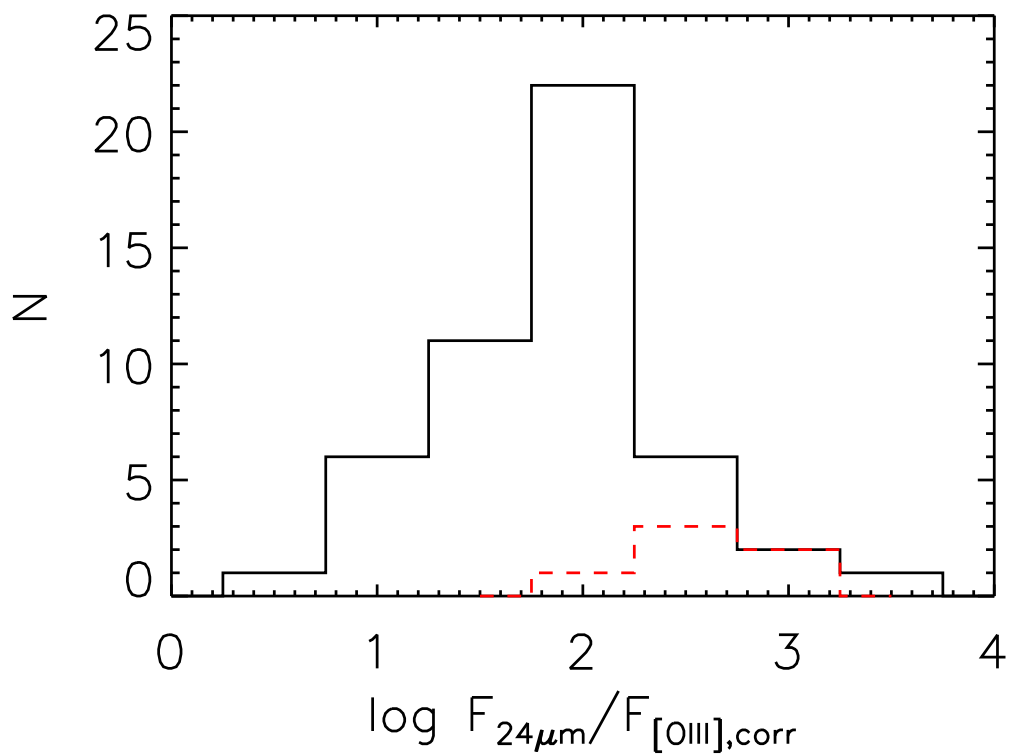,width=0.8\linewidth}
\caption{Histograms of the log of the ratio of the mid-IR (24$\mu$m) and extinction-corrected [O III]$\lambda$5007 emission-line luminosity for Type 2 AGN \protect\citep{lamassa10b} (black solid line) and the current sample of LBA-composites (red dashed line). The ratio is systematically higher in the LBA-composites ($2.71\pm0.34$) than that in Type 2 AGN ($1.89\pm0.57$), suggesting that the mid-IR continuum is primarily due to the starburst in most of the LBA-composites.\label{f:hist}}
\end{figure}

\end{document}